\documentclass{pasj00}
\draft

\begin{document}
\SetRunningHead{H. Matsuhara et al.}{Deep Extragalactic Surveys with AKARI (ASTRO-F)}
\Received{2005/11/14}
\Accepted{2006/04/28}

\title{Deep Extragalactic Surveys around the Ecliptic Poles with AKARI (ASTRO-F)}

%
%
\author{%
   Hideo \textsc{Matsuhara},\altaffilmark{1}
    \thanks{Further information contact Hideo Matsuhara (maruma@ir.isas.jaxa.jp)}
   Takehiko \textsc{Wada},\altaffilmark{1}
   Shuji \textsc{Matsuura},\altaffilmark{1} 
   Takao \textsc{Nakagawa},\altaffilmark{1} \\
   Mitsunobu \textsc{Kawada},\altaffilmark{2}
   Youichi \textsc{Ohyama},\altaffilmark{1}
   Chris P. \textsc{Pearson},\altaffilmark{1,11} 
   Shinki \textsc{Oyabu},\altaffilmark{1} \\
   Toshinobu \textsc{Takagi},\altaffilmark{1,3}
   Stephen \textsc{Serjeant},\altaffilmark{3,12}
   Glenn J. \textsc{White},\altaffilmark{3,12}
   Hitoshi \textsc{Hanami},\altaffilmark{4} \\
   Hidenori \textsc{Watarai},\altaffilmark{5} 
   Tsutomu T. \textsc{Takeuchi},\altaffilmark{6,13}
   Tadayuki \textsc{Kodama},\altaffilmark{7} 
   Nobuo \textsc{Arimoto},\altaffilmark{7} \\
   Sadanori \textsc{Okamura},\altaffilmark{8} 
   Hyung Mok \textsc{Lee},\altaffilmark{9} 
   Soojong \textsc{Pak},\altaffilmark{10} 
   Myung Shin \textsc{Im},\altaffilmark{9} \\
   Myung Gyoon \textsc{Lee},\altaffilmark{9} 
   Woojung \textsc{Kim},\altaffilmark{1}
   Woong-Seob \textsc{Jeong}, \altaffilmark{1}
   Koji \textsc{Imai},\altaffilmark{1} \\
   Naofumi \textsc{Fujishiro},\altaffilmark{1}\thanks{Present address : Genesia Corporation, Mitaka, Tokyo 181-0013}
   Mai \textsc{Shirahata},\altaffilmark{1} 
   Toyoaki \textsc{Suzuki},\altaffilmark{1} 
   Chiaki \textsc{Ihara}\altaffilmark{1}\thanks{Present address : NEC Aerospace systems, Ltd., 
Fuchu, Tokyo 183-8501} \\
   and
   Itsuki \textsc{Sakon}\altaffilmark{8} 
   }
 \altaffiltext{1}{Institute of Space and Astronautical Science, Japan Aerospace Exploration Agency, \\
   Sagamihara, Kanagawa 229 8510 } 
 \altaffiltext{2}{Graduate School of Science, Nagoya University, Chikusa-ku, Nagoya, 464-8602  }
 \altaffiltext{3}{University of Kent, Canterbury, Kent CT2 7NR, UK }
 \altaffiltext{4}{Iwate University, 3-18-8 Ueda, Morioka, 020-8550 } 
 \altaffiltext{5}{ALOS Project, Japan Aerospace Exploration Agency, Tsukuba, Ibaraki 305-8505 }
 \altaffiltext{6}{Laboratoire d'Astrophysique de Marseille, Traverse du Siphon, BP8 13376 \\
    Marseille Cedex 12, France}
 \altaffiltext{7}{National Astronomical Observatory of Japan, Mitaka, Tokyo 181-8588}
 \altaffiltext{8}{Department of Astronomy, School of Science, University of Tokyo, Bunkyo-ku, Tokyo 113-0033 }
 \altaffiltext{9}{Astronomy Program, Seoul National University, Shillim-Dong, Kwanak-Gu, Seoul 151-742, Korea}
 \altaffiltext{10}{Kyung Hee University, 1 Seocheon-dong, Giheung-gu, Yongin-si
    Gyeonggi-do 446-701, Korea} 
 \altaffiltext{11}{ISO Data Centre, ESA, Villafranca del Castillo, Madrid, Spain.}
 \altaffiltext{12}{Astrophysics Group, Department of Physics, The Open University, Milton Keynes, MK7 6AA, UK}
 \altaffiltext{13}{Astronomical Institute, Tohoku University, Aoba-ku, Sendai 980-8578 }


\maketitle

\begin{abstract}
AKARI (formerly ASTRO-F) is an infrared space telescope designed for an all-sky survey
at 10-180~$\mu$m, and deep pointed surveys of selected areas at 2-180~$\mu$m. The deep 
pointed surveys with AKARI will significantly advance our understanding
of galaxy evolution, the structure formation of the Universe,
the nature of the buried AGNs, and the cosmic infrared background. Here we describe 
the important characteristics of the AKARI mission: the orbit, and the attitude 
control system, and
investigate the optimum survey area based on the updated pre-flight sensitivities of 
AKARI, taking into account the cirrus confusion noise as well as the surface 
density of bright stars. The North Ecliptic Pole (NEP) is concluded to be the best 
area for 2-26~$\mu$m deep surveys, while the low-cirrus noise regions around the 
South Ecliptic Pole (SEP) are worth considering for 50-180~$\mu$m pointed 
surveys to high sensitivities limited by the galaxy confusion noise. Current 
observational plans of these pointed surveys are described in detail.
Comparing these surveys with the deep surveys with the Spitzer Space Telescope, the AKARI deep surveys 
are particularly unique in respect of their continuous wavelength coverage over the 2-26~$\mu$m range in 
broad-band deep imaging, and their slitless spectroscopy mode over the same wavelength range. 
\end{abstract}

\section{Introduction}\label{sec:introduction}

How did the galaxies form and evolve? How did the clusters of galaxies and the large-scale
structure of the Universe form and evolve? How do the Active Galactic Nuclei (AGN)
link together with the Ultralumonius Infrared Galaxies (ULIRGs) in their birth
and evolution? What is the nature of the sources contributing to the Cosmic Infrared 
Background (CIRB)? In order to answer these important questions of modern astronomy
we need statistically significant numbers of sources based on large-area 
surveys to uniform depths covering significant cosmological volumes.
Such surveys should be made in many wavebands, since
multi-wavelength spectral energy distributions (SEDs) as well as photometric redshifts
are useful to identify the types and ages of the detected galaxies.
  
Naturally, there is a trade-off, between the area and the depth of any survey, with
a greater required depth resulting in a narrower survey due to the finite
observation time. For example, at wavelengths shorter than  2~$\mu$m, towards two 
blank fields, the Hubble Space Telescope (HST) has produced excellent images to unprecedented sensitivity
and angular resolution (Hubble Deep Fields: HDF), but they were limited to relatively small areas 
(HDF-N, 5.3 arcmin$^2$, \cite{Will96} and HDF-S, 0.7 arcmin$^2$, \cite{Gad00}; 
\cite{Will00}). Recently, thanks to the {\it Advanced Camera 
for Surveys} onboard the HST, very large deep survey programs such as the Great Observatories Origins
Deep Survey (GOODS, 320~arcmin$^2$, \cite{Giav04}) and the Galaxy Evolution from Morphology
and SEDs survey (GEMS, 795~arcmin$^2$, \cite{Rix04}), have also been performed. A 1.5~deg$^2$ survey
program is now on-going (COSMOS project\footnote{See http://www.astro.caltech.edu/~cosmos/}).  

Longer infrared wavelengths beyond 2~$\mu$m are undoubtedly very important.
The stellar mass of a galaxy is a fundamental property of galaxies, and is well 
traced by the near-infrared ($1-2\,\mu$m) light. Since ground-based telescopes are not
sensitive enough to detect the stellar light beyond $2\,\mu$m, space-borne telescopes
are essential for the study of the galaxy stellar mass at high redshift.
Mid-IR wavelengths are also important because they are sensitive to dusty star forming 
galaxies which are characterized by the PAH features as well as the hot dust
continuum. This fact had previously been recognized by the first all-sky survey in the mid- and far-infrared 
with the {\it Infrared Astronomical Satellite} (IRAS, \cite{neu84}).  After the
launch of the {\it Infrared Space Observatory} (ISO, \cite{kes96}), which executed
 many infrared surveys covering a wide range in both depth and area at 7, 15,
90, and 170~$\mu$m,  various mid- to far-infrared SEDs of starburst galaxies as well
as AGN (\cite{GeCe00} and references therein) have been revealed and investigated to great detail.  The ISO surveys and their 
follow-up spectroscopic observations have demonstrated the importance of the mid- to far-infrared wavelengths 
in the study of evolution of galaxies
(HDF-S: \cite{fran03}, FIRBACK Marano: \cite{patr03}, ELAIS-S1/S2: \cite{Pozz04}, Lockman Hole: \cite{oya05}).

In August 2003, NASA launched an advanced space infrared observatory, the 
{\it Spitzer Space Telescope} ({\it Spitzer}; formerly SIRTF, \cite{wer04}).
 offering the capability of deep imaging from 3-160~$\mu$m and spectroscopy from 
5-38~$\mu$m.  In February 2006, JAXA (Japan Aerospace eXploration Agency)
launched a complementary infrared astronomy satellite AKARI 
(formerly known as ASTRO-F\footnote{ ASTRO-F was successfully launched on 21st February, 2006, 
and has been re-named {\it AKARI}.} ). AKARI will observe at near- to
far-infrared wavelengths like {\it Spitzer}.  It should be emphasized that AKARI is primarily 
a surveyor mission, though it has an additional role as an observatory with unique capabilities 
complementary to {\it Spitzer}.
A detailed comparison of these two missions are given in
section~\ref{subsec:comp}. It is expected that AKARI will also provide an invaluable 
database for the planning of observations with the  {\it Herschel} Space Observatory 
(\cite{pil03}).

In this paper, we describe the planned three major extragalactic deep surveys with
AKARI:
\begin{itemize}
\item a deep 2-26~$\mu$m survey at the North Ecliptic Pole (NEP), covering a 0.5~deg$^{2}$ area, to a
 5$\sigma$ flux limit of 10-30~$\mu$Jy at 11-18~$\mu$m, 
\item a wide, shallow 2-26~$\mu$m survey at the NEP, over a 6.2~deg$^{2}$ area, 
 to a 5$\sigma$ flux limit of 60-130~$\mu$Jy at 11-18~$\mu$m. 
  The survey region is a circular area surrounding the `` NEP-Deep" field, 
  and in total 450 pointing observations are required.
\item a deep 50-180~$\mu$m survey towards a low-cirrus region near the South 
  Ecliptic Pole (SEP) : covering 15~deg$^2$, down to the confusion limit due to unresolved galaxies.
\end{itemize}
The location of fields are summarized in Table~\ref{tab:epsvy}.
In section~\ref{sec:AKARI} we 
present the AKARI mission, and detail its observing modes. We then describe the 
key science objectives of the AKARI deep surveys in section~\ref{sec:objectives}. 
In section~\ref{sec:optimize} we discuss the various constraints on the survey characteristics, such as the visibility, 
confusion due to infrared cirrus and saturation from bright stars. 
Section~\ref{sec:plan} details the observing plan. Section~\ref{sec:uniq} compares 
these surveys with other infrared surveys with ISO and {\it Spitzer}.  
In section~\ref{sec:ground} we review the current status of our ground based 
preemptive observations.  The summary is given in section~\ref{sec:summary}. 
Throughout this paper we assume a flat concordance cosmology of $\Omega_{\rm M}=0.3$, $\Omega_{\rm \Lambda} = 0.7$
and $H_{\rm 0}=72 \,\rm km\,s^{-1}\, Mpc^{-1}$.

\begin{table}
  \caption{Overview of the planned AKARI extragalactic deep surveys}\label{tab:epsvy}
  \begin{center}
    \begin{tabular}{lcll}
	\hline \hline
Name     &  Field Center     & Field Size                       & Imaging Bands     \\
         &  (J2000)          & and Shape                        & and Depth         \\ \hline
NEP-Deep & \timeform{17h55m24.00s} \timeform{+66D37'32.0''} &  0.5~deg$^2$ circular & all IRC, FIS (Table~\ref{tab:nepdepth}) \\ 
NEP-Wide & \timeform{18h00m00.00s} \timeform{+66D36'00.0''} &  6.2~deg$^2$ circular & all IRC, FIS (Table~\ref{tab:nepdepth}) \\ 
SEP Low-Cirrus & \timeform{4h44m00.00s} \timeform{-53D20'00.0''} & 15~deg$^2$ fan-shape & S9W, S18W, all FIS (Table~\ref{tab:bandfl}) \\ \hline
    \end{tabular}
  \end{center}  
\end{table}

\section{The AKARI Mission Outline}\label{sec:AKARI}

AKARI will be a second
generation infrared sky survey mission after IRAS with much improved sensitivity, 
spatial resolution and wider wavelength coverage. During the expected mission 
life of more than 500 days, AKARI will make an all-sky survey in 6 
wavebands in the mid- to far-infrared since IRAS, and the first ever all-sky
 survey at 100-160~$\mu$m. In addition to the all-sky survey, deep imaging and spectroscopic observations in the pointing mode will also
be carried out in many wavebands covering 2-160~$\mu$m. The observation strategy
in this mode was discussed in \citet{peama01} based on the preliminary mission specifications.
This paper describes the deep extragalactic surveys, based on the
updated pre-flight specifications of the mission.  The expected results of the deep pointed surveys will appear in an updated version of the above paper based on the in-flight sensitivities (Pearson et al. 2006, in preparation). 

AKARI is equipped with a 68.5-cm cooled (approximately to 6K) Ritchey-Chr\'{e}tien 
type telescope and two focal plane instruments covering the 2-180~$\mu$m wavelength
range. Details of the AKARI mission and satellite system are described in \citet{mur04},
 and details on the current performance of the focal plane instruments are
described in \citet{kaw04} and \citet{ona04}. 

\subsection{AKARI orbit and observation modes}

\begin{figure}
  \begin{center}
    \FigureFile(150mm,75mm){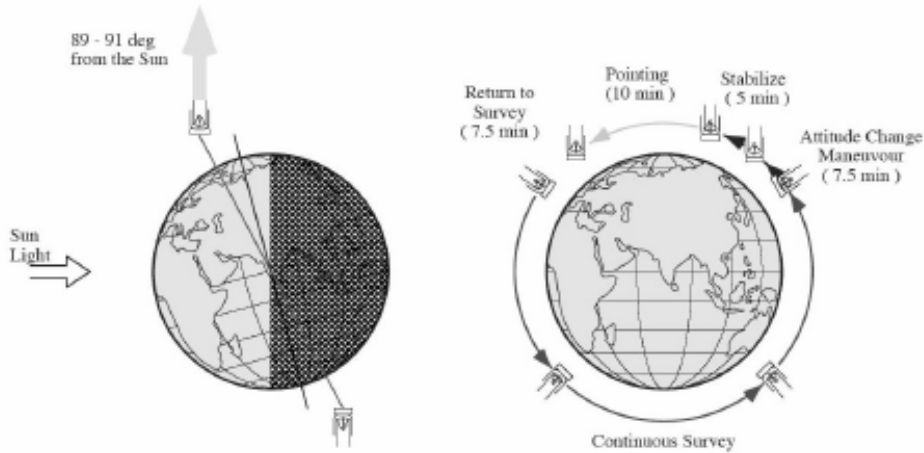}
  \end{center}
  \caption{The AKARI spacecraft attitude is shown schematically. 
(Left) the all-sky survey mode.
(Right) the attitude operation for the pointed observations. The duration of 
a single pointed observation is limited to approximately 10 minutes. The pointing direction is
restricted to within $\pm 1$~deg in the direction perpendicular to the orbital
plane.}\label{orbit}
\end{figure}

The orbit of AKARI is a sun-synchronous polar orbit at an altitude of 750km and an
inclination of 98.4 deg. The orbital period is approximately 100~minutes. In Figure~\ref{orbit}, the AKARI spacecraft attitude is 
shown schematically. In the all-sky survey mode, the spacecraft rotates uniformly 
around the axis directed toward the Sun once every orbital revolution, resulting 
in a continuous  scan of the sky on one orbit. In this configuration, the entire sky can be covered in half a year.
In the attitude operation for pointed observations (Figure~\ref{orbit} right),
due to the limitations imposed to prevent earthshine from illuminating the telescope baffle, 
the duration of any single pointed observation is limited to 10 minutes. The pointing direction can 
be freely chosen in the telescope orbital plane given by the survey mode attitude. 
However, it is restricted to within $\pm 1$~deg to the direction perpendicular to the 
orbital plane\footnote{We hereafter call this function of the attitude control system
``offset control" which allows the telescope to be directed out of the nominal 
orbital plane up to $\pm 1$~deg.}. The attitude control system provides an additional
capability to slightly change the pointing direction defining the following attitude modes:
step-scan, slow-scan, and micro-scan.  The slow-scan is a uniform scan with much 
slower scan speed (less than 30 arcseconds per second) than that of the all-sky
survey, and is an option for pointed observations with AKARI. Hence the duration of this observing
mode is also limited to 10~minutes. The slow-scan survey is especially useful for covering a larger sky 
area than that obtainable by pointing toward the fixed direction in the sky to much better 
sensitivities than those of the all-sky survey.
One should note that the overheads  in the pointing attitude mode are significant: 
approximately 20 minutes is required for the maneuvering and stabilization. Hence 
the number of pointing observations in one orbit is restricted to three or less (see
Table~\ref{tab:orbit}).

\begin{table}
  \caption{Parameters of the AKARI Orbit and Operations}\label{tab:orbit}
  \begin{center}
    \begin{tabular}{ll}
	\hline \hline
	Orbit Parameters        &               \\
	\hline
	Altitude                                & 750 km        \\
	Period                                  & 99.83 minutes \\
	Mean motion of Argument of Perigee     & 0.9856 deg $\rm day^{-1}$ \\
	\hline
	Operational Parameters  &               \\
	\hline
	Maximum offset angle out of the observing plane & $\pm 1$~deg \\
	Roll angle offset & not used \\
	Maneuver time from the survey mode to the pointing mode & 450 sec \\
	Attitude stabilization time before the pointed observation & 300 sec max.	 \\
	Duration of a pointed observation & 600 sec nominal \\
	Maneuver time from the pointing mode to the survey mode & 450 sec \\
	\hline
    \end{tabular}
  \end{center}
\end{table}

\subsection{Scientific instruments}

\begin{figure}[h]
  \begin{center}
    \FigureFile(150mm,80mm){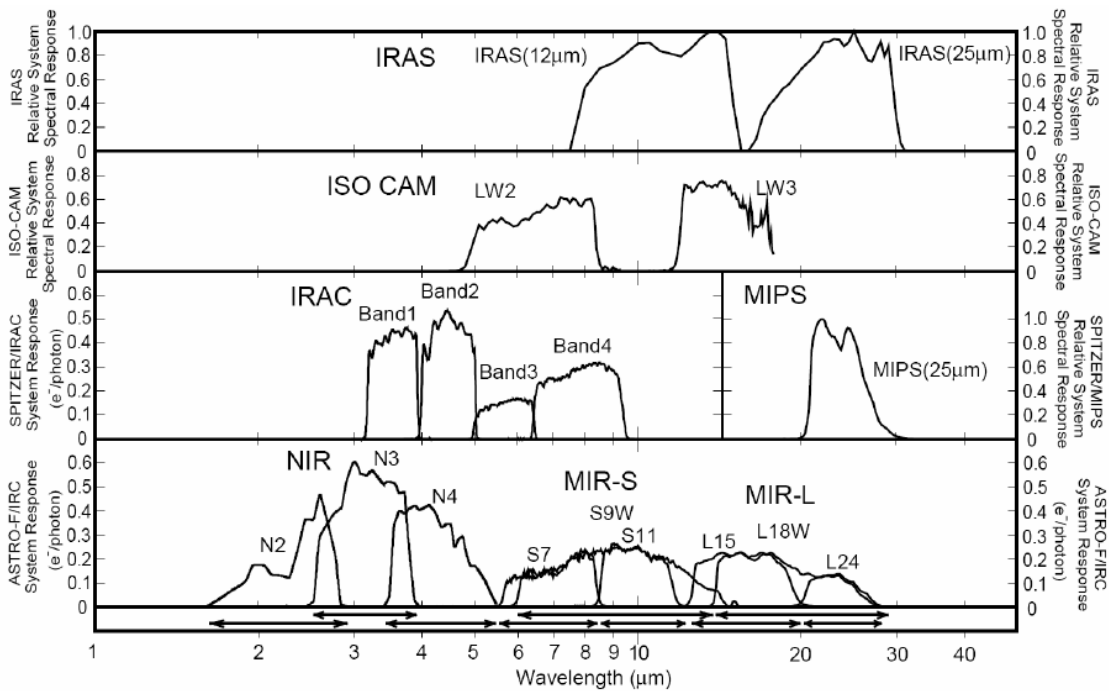}
  \end{center}
  \caption{Comparison of spectral responce curves of AKARI wavebands for imaging and survey with those of other 
infrared astronomical satellites. Those of ISO and IRAS are relative, while those of {\it Spitzer} and AKARI are
absolute in units of electrons per photon.
}\label{specres}
\end{figure}

AKARI incorporates two focal plane instruments covering the infrared wavelength range from
2 to 160~$\mu$m. The first focal plane instrument is the Far-Infrared Surveyor 
(FIS, \cite{kaw04}) which will primarily survey the entire sky simultaneously in 
four far-infrared bands from 50 to 160~$\mu$m with approximately
diffraction-limited spatial resolutions (30-60 arcsec, \cite{jeon03}). Each waveband is equipped with
corresponding far-infrared imaging arrays denoted as: N60, WIDE-S, WIDE-L, and N160, all of which
are operated simultaneously.
The second focal-plane instrument is the InfraRed Camera (IRC, \cite{ona04}), 
which is primarily designed to take deep images of selected regions of sky 
by pointed observations from 2 to 26~$\mu$m. The IRC comprises of three
channels: NIR(2-5~$\mu$m), MIR-S(5-12~$\mu$m), MIR-L(12-26~$\mu$m). These three channels
can be operated simultaneously. The waveband
characteristics for broad-band imaging are summarized in Table~\ref{tab:bandfl} and Figure~\ref{specres}
Each IRC channel is equipped with 3 broad-band imaging bands, which have to be selected
by rotating a filter wheel.
Both focal plane instruments are equipped with spectroscopic elements of 
moderate or low resolution : the FIS is equipped with a Fourier-Transform Spectrometer
(FTS) covering 50-180~$\mu$m using the WIDE-S and WIDE-L detector arrays, while the IRC is
equipped with low-dispersion grisms (and a prism in the NIR channel) which enable slitless spectroscopy over the field-of-view (FOV) of each channel with spectral resolutions of
40-90 at 2-26~$\mu$m. The IRC also has the capability for slit spectroscopy.

\begin{figure}[h]
  \begin{center}
    \FigureFile(80mm,80mm){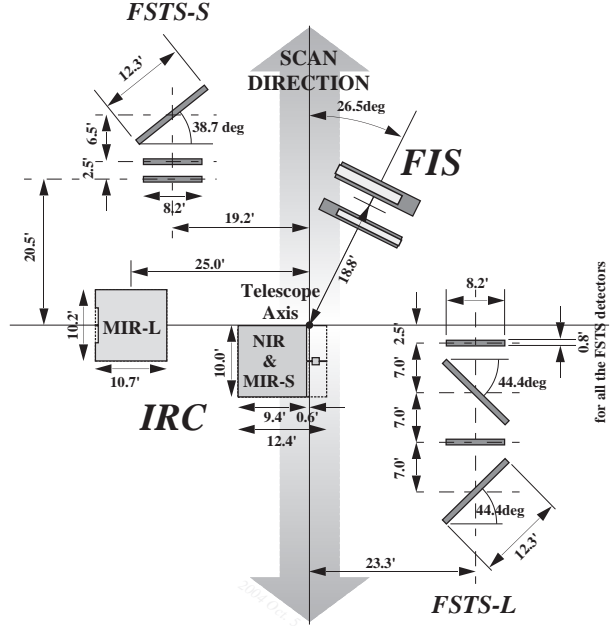}
  \end{center}
  \caption{Configuration of the field-of-views of the AKARI focal-plane instruments,
projected onto the sky.}\label{fov}
\end{figure}

Figure~\ref{fov} shows the FOV configuration projected onto the sky. Note that the NIR and the MIR-S
channels observe the same sky simultaneously, while the FOV of the MIR-L channel is
offset by approximately 25 arcmin from the NIR and the MIR-S channels.  During the pointing observations, all focal plane 
instruments are operated simultaneously although the target is acquired through a selected FOV
of either the NIR/MIR-S, MIR-L, or  FIS. Hence parallel-mode observation, {\it i.e.} 
observation with channels which do not observe the same field of view in the sky, is always executed as long as
this does not have any impact on the data rate or does not induce heat dissipation to the AKARI
cryostat, which would shorten the mission life.  
The pre-flight sensitivities estimated from the current laboratory experiments of the
telescope and the focal plane instruments at low temperatures are shown in 
Tables~\ref{tab:bandfl} and \ref{tab:specspec}. The IRC sensitivity is given for its single-filter astronomical
observation template (AOT), in which no filter change is made during a single pointed
observation, such that each IRC channel observes the target with a single filter. There are 
also two-filter or three-filter AOTs as well as an AOT dedicated for the spectroscopy.
The FIS broad-band sensitivity in pointing
mode is not available since only the FTS will be used in the pointing observation mode. 
The FIS sensitivity in the slow-scan attitude mode is shown, since 
the slow-scan survey is useful to observe larger areas within the limited time of a single
pointed observation. 

\begin{table}
  \caption{Specifications of IRC and FIS filter bands and the flux limits in various observing modes}\label{tab:bandfl}
  \begin{center}
    \begin{tabular}{llllll}
	\hline \hline
Band     &  $\lambda_{\rm c}$ & $\Delta \lambda$  & All-sky survey    & Slow-Scan(15"sec$^{-1}$)  & Pointing\footnotemark[$*$] \\
(IRC)    &  ($\mu$m)          & ($\mu$m)          & (5$\sigma$, mJy)  & (5$\sigma$, mJy)          & (5$\sigma$, $\mu$Jy) \\ \hline
N2       &    2.43            &   0.68            &    N/A            &    N/A                    &    7.9               \\ 
N3       &    3.16            &   1.12            &    N/A            &    N/A                    &    3.7               \\ 
N4       &    4.14            &   1.22            &    N/A            &    N/A                    &    7.2               \\ 
S7       &    7.3             &   2.6             &    N/A            &    11                     &     33               \\ 
S9W	  &    9.1             &   4.3             &    80             &     7                     &     26               \\ 
S11      &    10.7            &   4.7             &    N/A            &    13                     &     37               \\ 
L15      &    15.7            &   6.2             &    N/A            &    16                     &     68               \\ 
L18W	  &     18.3           &   10.0             &    130            &    20                     &     87               \\ 
L24      &    23.0            &   5.4             &    N/A            &    53                     &    180               \\ \hline
Band     &  $\lambda_{\rm c}$ & $\Delta \lambda$  & All-sky survey\footnotemark[$\dagger$]    & Slow-Scan(15"sec$^{-1}$)\footnotemark[$\dagger$]  & Pointing  \\
(FIS)    &  ($\mu$m)          & ($\mu$m)          & (5$\sigma$, mJy)  & (5$\sigma$, mJy)          & (5$\sigma$, $\mu$Jy) \\ \hline
N60      &    65              &   21.7            &   1000            &    70                     &    N/A                \\
WIDE-S   &    90              &   37.9            &    200            &    14                     &    N/A                \\
WIDE-L   &    140             &   52.4            &    400            &     9                     &    N/A                \\
N160     &    160             &   34.1            &    800            &    18                     &    N/A                \\ \hline
         &                    &                   &                   &                           &                       \\
\multicolumn{6}{l}{\parbox{130mm}{\footnotesize \par \noindent
 \footnotemark[$*$] in case of the astronomical observation template (AOT) for deep imaging with no  
    filter change and no dithers at Ecliptic poles.  Net exposure times in this AOT are 459~sec
    per pointing for the NIR, and 513~sec per pointing for the MIR-S/MIR-L channels, respectively.
    Photometric apertures are determined so that the signal-to-noise ratio for the source is
    maximized with the expected point-spread function at each waveband (typical aperture radius is 1-2 pixels). 
    \par \noindent
 \footnotemark[$\dagger$] sensitivity for one scan. Surveys with at least two scans (all-sky survey) or four
   scans (SEP low-cirrus region survey) are planned.
}}
  \end{tabular}
  \end{center}  
\end{table}

\begin{table}
  \caption{Specifications of IRC spectroscopic channels}\label{tab:specspec}
  \begin{center}
    \begin{tabular}{llll}
	\hline \hline
Band     &  Wavelength Coverage & Spectral Resolution      & Flux limit (continuum) \footnotemark[$*$] \\
         &  ($\mu$m)            & ($\lambda / d\lambda$) \footnotemark[$\dagger$] & (per pixel, 5$\sigma$, mJy) \\ \hline
NP       &    1.7 -- 5.5        &      22                  &      0.02                                 \\ 
SG1      &    5.5 -- 8.3        &      47                  &      1.0                                  \\ 
SG2      &    7.4 -- 13.        &      34                  &      3.2                                  \\ 
LG2      &    17.7 -- 25.       &      27                  &      6.7                                  \\ \hline
\multicolumn{4}{l}{\parbox{100mm}{\footnotesize \par \par \noindent
  \footnotemark[$*$] in case of one pointing observation at ecliptic poles with the AOT for the ``slitless spectroscopic
  mode", where the source spectra are dispersed in the imaging area of the detector array.
  \par \noindent
  \footnotemark[$\dagger$]  $\lambda$ is the center wavelength, and $d\lambda$ is a wavelength resolution corresponding 
  to the FWHM size of the point spread function.
 }} 
    \end{tabular}
  \end{center}  
\end{table}

\subsection{Mission phases and observation programs}

Observations with AKARI can be divided into the following phases:
\begin{itemize}
\item{PV phase [ L(launch)+1 -- L+2months ] : observations for performance 
verification of the whole system.}
\item{Phase 1 [ L+2 -- L+8months ] : primarily for the all-sky survey. 
However, pointed observations toward fields close to the ecliptic poles are planned for 
orbits passing through the South Altantic Anormaly (SAA).  }
\item{Phase 2 [ L+8 -- L+18months (boil-off of liquid He) ] : supplemental 
all-sky survey observations for areas not sufficiently covered in phase-1, and pointed 
observations in any region on the sky.}
\item{Phase 3 [after boil-off of liquid He ] : pointed observations with 
the IRC-NIR channel which will continue to function even after the boiling-off of the 
liquid He. The duration of this phase
  is uncertain and is limited by the life of the mechanical coolers. }
\end{itemize}

During the AKARI mission phases above, the following
observation programs are planned:

\begin{itemize}
\item{Large-area Survey (LS): the all-sky survey, the primary objective 
of the AKARI mission, as well as large-area {\it Legacy} surveys in pointing
mode over two dedicated fields near the ecliptic poles: the NEP (section~\ref{subsec:nep})
and the Large Magellanic Could (LMC). The outline of the LS programs are described in \citet{maruma05}. }
\item{Mission Programs (MP): an organized program of pointed observations 
planned by the AKARI science working groups (AKARI team plus collaborators 
within the international astronomical community).
The targets are spread over large areas but cannot be included in the NEP and 
LMC surveys.}
\item{Open-time Program (OP): a program of pointed observations open to the Japanese, Korean, and ESA communities. 30\% of 
the total pointed observation opportunities in both phases 2 and 3 will be allocated to OP. }
\item{Director's Time: observing time allocated to the Project Manager, 
including any ``Target of Opportunity" time.
Less than 5\% of the total observation is  considered so far.}
\item{Calibration Time: for the long-term maintenance and performance 
verification for each instrument, calibration observations will be made 
at regular periods.}
\end{itemize}

\section{Key Scientific Objectives}\label{sec:objectives}

Here we describe the key scientific objectives of the AKARI deep pointed observation extragalactic 
surveys, and also discuss the optimal survey area and depth. 

\subsection{Mass assembly and structure evolution}\label{sec:variance}

\begin{figure}[ht]
  \begin{center}
    \FigureFile(100mm,110mm){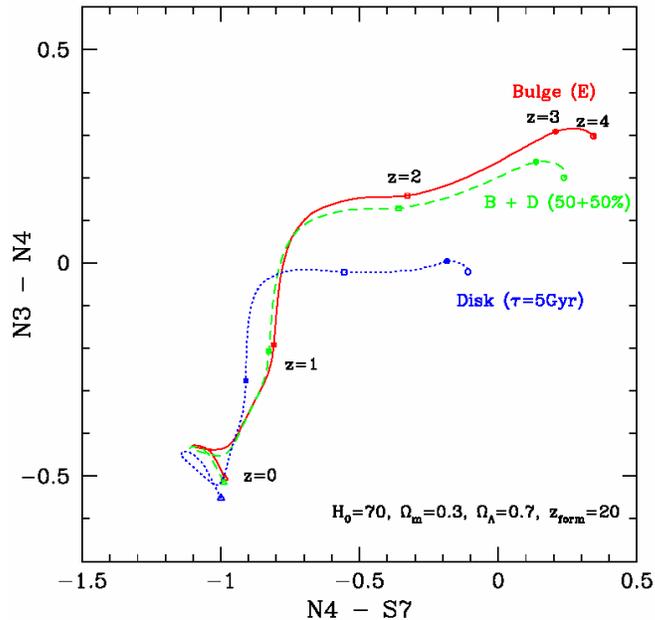}
  \end{center}
  \caption{Tracks of early and late type galaxies on the two colour 
	([N4] - [S7] vs [N3] - [N4] in AB magnitudes) diagram. The SEDs are calculated from the
	model of \citet{KA97}.
  }\label{klmirc} 
\end{figure}

\begin{figure}[ht]
  \begin{center}
    \FigureFile(100mm,110mm){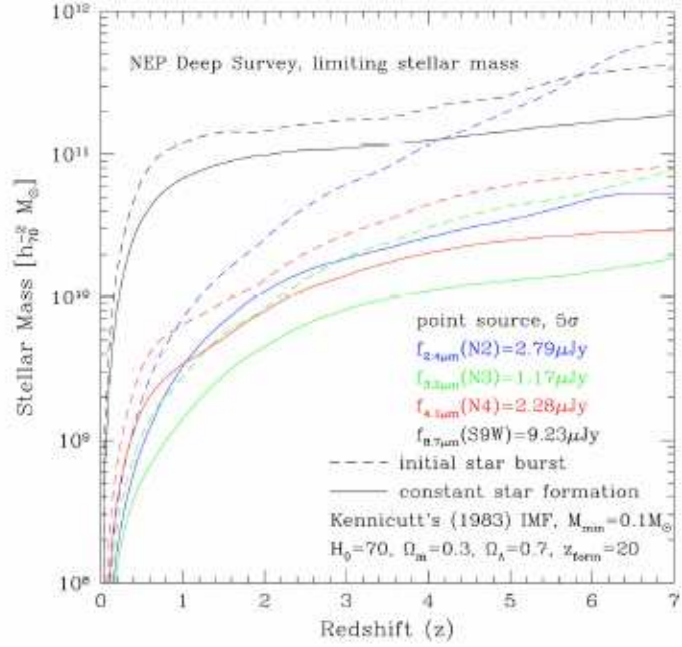}
  \end{center}
  \caption{Minimum detectable stellar mass of galaxies as a function
	of redshift for 8(N2), 10(N3, N4) pointing observations per filter with 
	AKARI/IRC. Two extreme cases of the star-formation history are assumed:
	an initial burst (dashed lines) and constant star-formation (solid lines) to form the stellar mass
 	at the detection limit during the age of the galaxy .
  }\label{massvsz} 
\end{figure}

The total stellar mass of a galaxy is well traced by the near-infrared 
luminosity at 1-2~$\mu$m where the effects of star-formation are small compared
to optical wavelengths. The evolution of the stellar mass function 
of the galaxy provides a basic piece of information for modern 
cosmology. Moreover, the galaxy's stellar mass can be used as a tool 
to reveal the evolution of the large-scale structure of the Universe, if a sufficient
number of galaxies are sampled over a large enough cosmic volume.
Since the spectroscopic determination of the redshift of galaxies in such a large  
sample is extremely time and effort consuming, the ``photo-$z$" technique, a method of the determination
of redshift based on multi-wavelength data, has been widely used (see {\it Hyper-z}: \cite{bolz00},
Baysian method: \cite{beni00}, {\it ANN-z}: \cite{collis}), \cite{dick03}
(HDF-N), \cite{font04} (K20 survey) for examples).
Since ground-based telescopes are not sensitive enough to detect the 
stellar light beyond 2~$\mu$m, the photo-$z$ technique has been thus far based on the 0.4~$\mu$m break and the Lyman break.
On the other hand, space-borne telescopes, such as {\it Spitzer} and AKARI, can observe
longward of 2~$\mu$m and can thus execute a systematic study 
of the galaxy stellar mass at high redshifts. Furthermore, there is another
redshift estimator in the rest-frame 1.6~$\mu$m where the $\rm H^{-}$ opacity minimum 
of the stellar atmosphere \citep{saw02} exists. The 4 
{\it Spitzer}/IRAC wavebands are specially designed to determine the photometric redshift
by using this bump. The near-infrared and shorter mid-infrared filter bands of IRC 
(see Table~\ref{tab:bandfl}) continuously cover the $2-10\,\mu$m wavelength range and hence 
a systematic study of the galaxy mass at various redshifts is also possible. 
Figure~\ref{klmirc} shows the tracks of typical normal galaxies for $z$=0-4 on 
a two colour diagram for the IRC wavebands. For the determination of the mass 
function in each redshift bin, a mass sensitivity to well below 
$10^{11}$~M$_{\odot}$ ({\it i.e.} charanteristic stellar mass $M_{*}$ of the Schechter function, see \cite{dick03})
 is required. As shown in Figure~\ref{massvsz}, 10 pointing 
observations in the N3 and N4 bands satisfy this requirement out to $z$=4.
In this plot $z_{form}=20$ is assumed to demonstrate the high sensitivity to the 
stellar mass even at high redshifts.

\begin{figure}[htb]
  \begin{center}
    \FigureFile(100mm,100mm){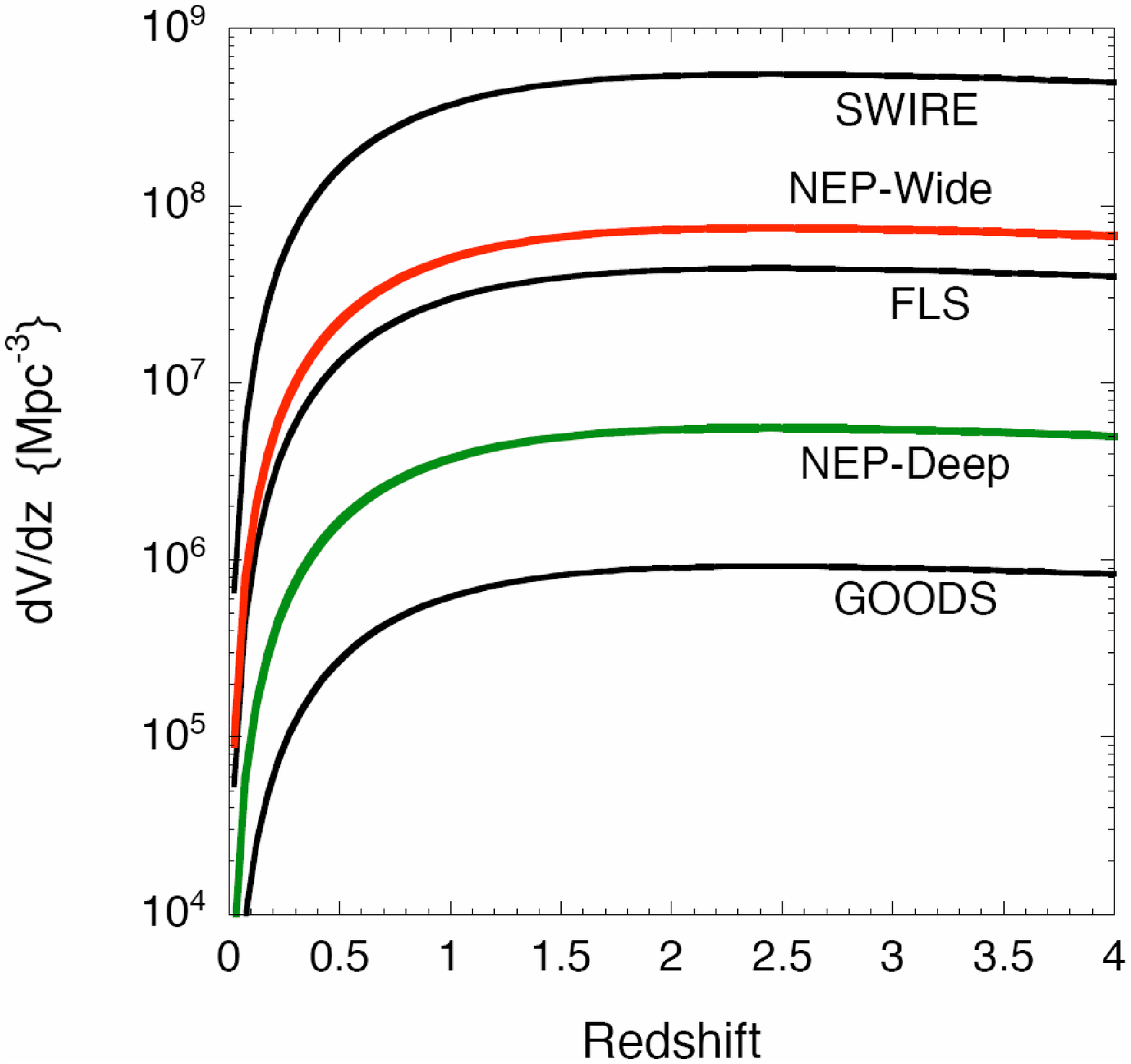}
  \end{center}
  \caption{Comparison of the comoving volume per unit redshift for {\it Spitzer}/GOODS, SWIRE
	and proposed AKARI deep surveys(``NEP-Deep" and ``NEP-Wide").
  }\label{comvol}
\end{figure}

Here we discuss the requirement for the survey volume.
We require statistically significant numbers of galaxies detectable 
with AKARI at each redshift bin. If the source of uncertainty is solely Poisson
error, one could simply request $\ge 100$ sources per redshift bin with 10\% accuracy.
However, for relatively small survey volumes, observational estimates of the number
density of the galaxies suffer from the cosmic variance - the field-to-field variation
due to the large scale structure. For example, in case of a 0.5~deg$^{2}$ area, the   
comoving volume for $\Delta z$=0.2 intervals at $z$=2-3 is $1.1 \times 
10^{6}$~Mpc$^{3}$, corresponding to a cube of side length of 100Mpc. According to 
\citet{som04}, the uncertainty due to the cosmic variance for strongly 
clustered populations such as EROs is estimated to be 30-40\%, while for less clustered 
populations it is around 15-20\%(for LBGs). The cosmic variance shows a rather weak dependence 
on the survey volume $V$: $V^{\gamma /6}$ where $\gamma$ is the index of the two-point
correlation function, and therefore a factor of two reduction of the variance requires
an order of magnitude larger volume which cannot be easily obtained by the 
on-going space infrared telescopes. It should be noted however, that \citet{pedo94} 
described the power spectra obtained by optical large-area surveys 
as well as the IRAS survey which gives only 3-4\% cosmic variances for 0.5~deg$^{2}$ 
area with $\Delta z$=0.2 interval.  In conclusion at $z \geq 1$ a 0.5~deg$^{2}$ area
is our minimum requirement to avoid serious effects due to cosmic variance, while 
for $z \leq 1$, more than 5~deg$^{2}$ area is recommended.

The survey area and depth requirements to obtain the desired scientific objectives described
in this subsection are summarized as follows:
\begin{itemize}
\item survey depth: sensitivity for the stellar mass of a galaxy: $<10^{11}$~M$_{\odot}$ out 
  to $z$=4. Approximately 10 pointing observations are required for the 2-9~$\mu$m
  wavebands.
\item survea area : 0.1~deg$^2$ is not suffcient to overcome the cosmic varience,
  however, 0.5-1~deg$^2$ area is reasonable at $z=1-4$ if a 10-30\% cosmic variance 
 effect is acceptable.   
\end{itemize}

\subsection{Dusty star-formation history of the universe}\label{sec:dusty}

In the past few years considerable progress has been made in the understanding
of the star formation rate of high-$z$ galaxies using the Lyman drop out 
technique to identify galaxies at $z=3-5$ (\cite{ste96}; \cite{ouc04}). 
Well over 2000 Lyman-break systems have now been observed at $z>3$. These are 
ultraviolet (UV)-bright, actively star-forming galaxies with blue spectra and star 
formation rates of a few to $\sim 50 \, \rm M_{\odot} yr^{-1}$.
Moreover, thanks to the commissioning of very deep large-area surveys with 
narrow-band filters on 8-m class telescopes, more than 
twenty Ly$\alpha$-emitting(LAE) galaxies at $z > 5$ have been identified (\cite{tan03}, \cite{shi04},
\cite{ouc05}), providing us with the first insight into the star formation history of the Universe beyond 
$z$=5. Typical star formation rates (SFR) of $z=5-6$ LAEs are 
$5 - 10\, \rm M_{\odot} yr^{-1}$, which however should be considered as a lower 
limit, since a blue half of the Ly$\alpha$ emission may be absorbed
by HI gas and dust grains within the galaxy itself, and by the intergalactic HI gas. 

The popular Madau plot \citep{mad96}  implied
that the SFR of the Universe peaked at $z=1-2$, although the SFR derived from the rest-frame 
optical/UV light has serious uncertainties in the correction of the dust extinction. However, the stellar
mass of a galaxy derived from SED fitting including the $K$-band does not suffer so critically from the extinction
corection, and it has been revealed that stellar mass density of the Universe as a 
function of redshift shows a rapid increase in the stellar mass density at $z=1-2$,
 based on deep and wide $K_{\rm s}$ band surveys (\cite{capu04}, \yearcite{capu05}). 
Furthermore, galaxies selected from optical and near infrared photometry with 
$BzK = (z'-K)_{\rm AB} - (B-z')_{\rm AB} > -0.2$  have been found to show 
active star formation rates of $\rm > 100 \, M_{\odot} yr^{-1}$ at $z > 1.4$ 
\citep{dad04}. Therefore, it is very important to understand the star 
formation history of the Universe not only at $z > 4$ but also at $z=1-2$.
 
\begin{figure}[ht]
  \begin{center}
    \FigureFile(150mm,130mm){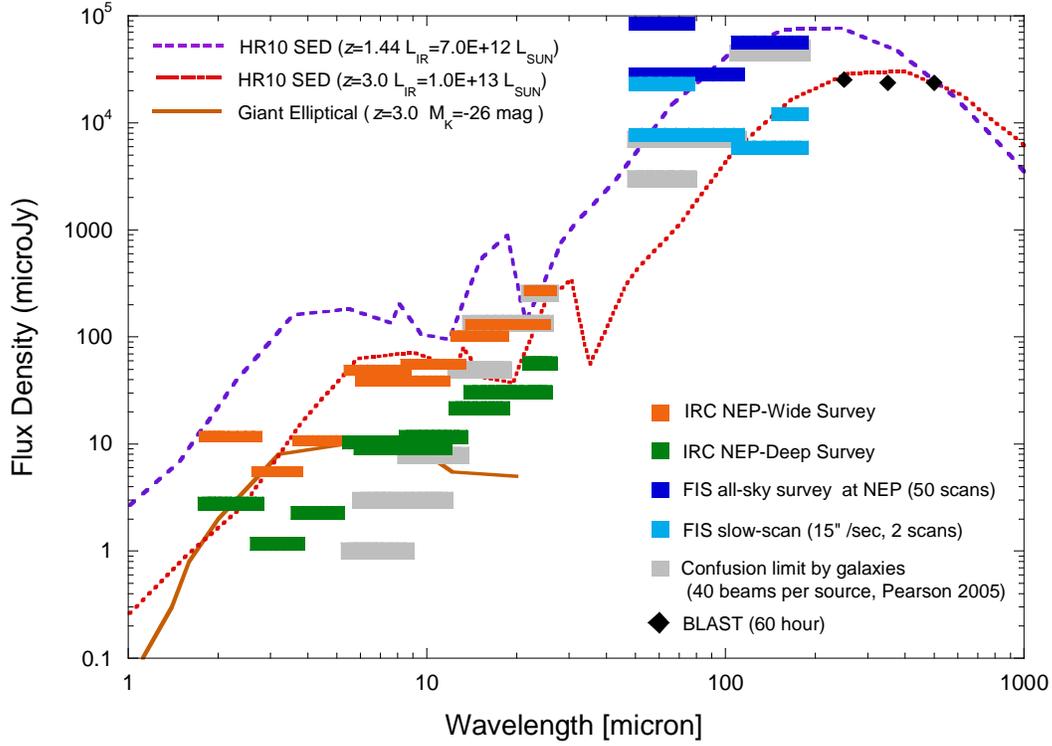}
  \end{center}
  \caption{5$\sigma$ sensitivities in micro Jansky for the IRC instrument in pointing mode 
	(``Deep" and ``Wide" surveys, see section~\ref{sec:plan}) and
	the FIS instrument in the all-sky 
	survey mode as well as in the slow-scan mode.  Also shown 
	for reference are the assumed spectral energy distributions of a normal 
	spiral galaxy at $z$=3 (rest-frame absolute K = $-26$~mag (in Vega)) and dusty ERO spectral 
	energy distributions  at redshifts of 1.44 and 3 modeled on the 
	archetypical ERO HR10. The source confusion due to galaxies have been included 
	from the recently updated evolutionary models of \citet{pea05}.
  }\label{sed}
\end{figure}

It has been revealed that a significant portion of the energy emitted 
in the early Universe came not only from UV-selected galaxies but also from 
very luminous galaxies in the far infrared (FIR), that are hidden at optical 
wavelengths because of obscuration from interstellar dust grains. Recent submillimeter deep 
surveys with the SCUBA instrument on the JCMT have shown the existence of 
numerous such sources, showing that the sub-mm galaxies represent
an important component of the cosmic star formation at high redshifts. 
For example, from the follow-up optical spectroscopy with multi-object 
spectrographs on 10-m class telescopes. \authorcite{chap03} (\yearcite{chap03}, \yearcite{chap05})
have reported a median redshift of 2.3 for their sample of sub-mm galaxies with  VLA identifications.

Since the submillimeter galaxies are very faint at optical wavelengths
(UV in their rest frame) due to extinction by dust in the systems themselves, mid- and 
far-infrared observations with space infrared telescopes such as {\it Spitzer} and 
AKARI are critical to unveil their true nature. Such infrared observations will 
give unique information on their infrared SEDs which is an important clue to 
obtain information on the mass-to-light ratio, the overall dust opacity, and
to infer the hidden star formation rate of these galaxies.

\begin{figure}[htbp]
  \begin{center}
    \FigureFile(160mm,240mm){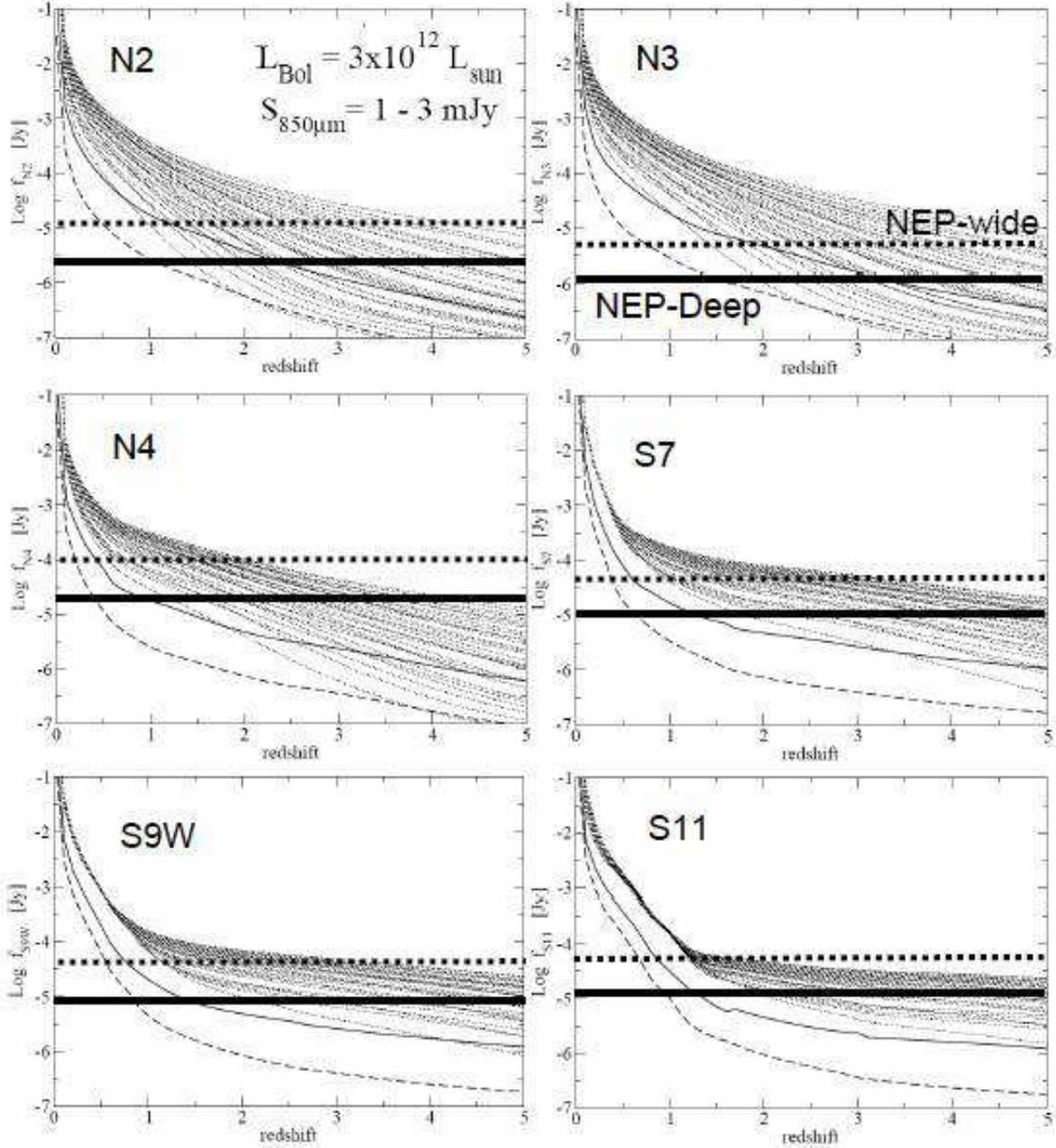}
  \end{center}
  \caption{Predicted fluxes of ULIRGs as a function of redshift for the IRC-NIR and IRC-MIR-S bands.
	The SED variation is determined from the observed 
	SED of submm galaxies at $z=2-3$ \citep{takpea05}. The luminosity 
	is assumed to be $3 \times 10^{12}\, L_{\odot}$. Thick solid and dashed lines
	indicate the expected fluxes from Arp 220 and M82. The horizontal lines 
	indicate the flux limits of the NEP survey with the IRC (dashed : for ``NEP-Wide", solid : for
	``NEP-Deep").
 }\label{fluxvsza}
\end{figure}

\begin{figure}[htbp]
  \begin{center}
    \FigureFile(160mm,240mm){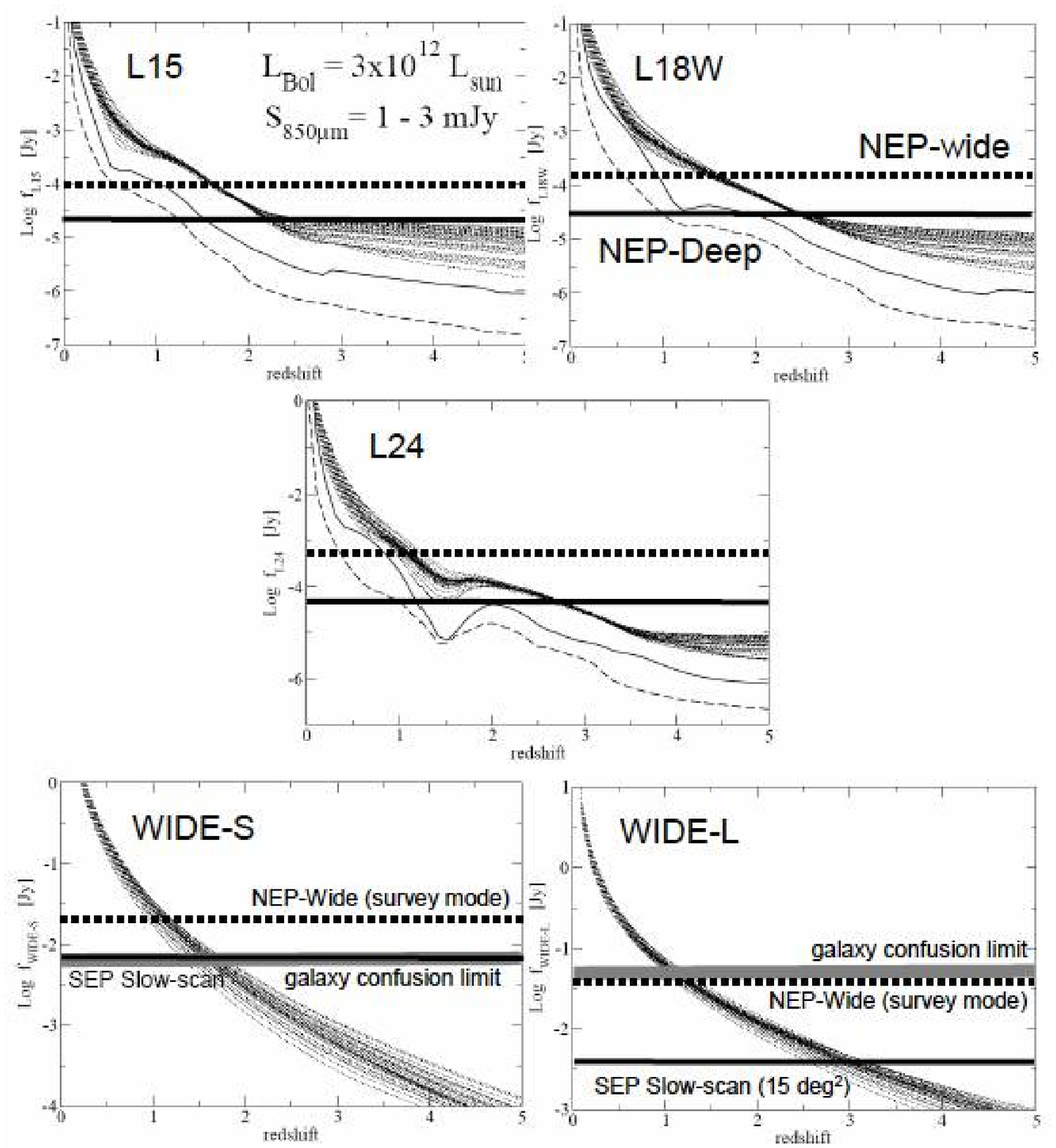}
  \end{center}
  \caption{Same as Figure~\ref{fluxvsza}, but for the IRC-MIR-L and the FIS bands.
 }\label{fluxvszb}
\end{figure}

From these consideration, we propose to concentrate on the dusty
star formation history at $z$=1-3 with AKARI, and generate statistically
meaningful numbers of samples (order of 100-1000) in each redshift bin.
Figure~\ref{sed} shows the expected flux limits for the AKARI wavebands 
as well as the SED of a hyper-luminous ($L_{\rm IR} = 10^{13}\,L_{\odot}$) 
ERO at $z$=3. Quantitative discussion on the expected number
 of sources in the far infrared was given in \citet{jeon04}, and its revision
including the near and mid infrared source counts can be found in Pearson et al. (2006) (in preparation). 
Figures~\ref{fluxvsza}, \ref{fluxvszb} show predicted fluxes of ULIRGs
as a function of redshift for the IRC and FIS wavebands used for the proposed survey
toward the ecliptic poles. Fluxes for various ULIRG SEDs
 with infrared luminosity of $3 \times 10^{12}\, L_{\odot}$ \citep{takpea05}
 are shown. Note that if there are AGN contributions, the ULIRGs should be brighter than that 
expected from this model. Hence in order to detect ULIRGs at $z$=2-3 in the mid infrared,
8-10 pointing per filter, {\it i.e.} the depth proposed for the ``NEP-Deep" survey,
is sufficient.  It should also be noted that the FIS slow-scan survey can potentially 
 detect the far-infrared emission of the $z\sim2$ mid-infrared sources 
in the ``NEP-Deep" field (Figure~\ref{fluxvszb}).

Figure~\ref{comvol} shows a comparison of the comoving volume per unit redshift 
interval for {\it Spitzer}/GOODS (160 arcmin$^{2}$ for CDF-S, \cite{Giav04}, 
{\it Spitzer}/SWIRE(50~deg$^{2}$, \cite{lons04}), and for the proposed AKARI
NEP surveys. As discussed in section~\ref{sec:variance}, a 0.5~deg$^2$ survey
provides the sufficient cosmic volume for $z \geq 1$.

 The survey area and depth requirements to obtain the desired scientific objectives described
in this subsection are summarized as follows:
\begin{itemize}
\item survey depth: the sensitivity to detect ULIRGs at high redshift, for example, 
($L_{\rm IR} = 10^{13}\,L_{\odot}$$<10^{11}$~M$_{\odot}$ out to $z$=3. 
In particular, the 11-24~$\mu$m wavebands are important, and approximately 10 pointing
 observations are required in each band.
\item survea area : 0.1~deg$^2$ is not suffcient to overcome the cosmic variance,
  however, a 0.5-1~deg$^2$ area is reasonable for $z=1-3$ if 10-30\% cosmic varience 
  can be acceptable.  Below $z<1$ a 5-10~deg$^2$ area is preferable. 
\end{itemize}


\subsection{Unveiling the nature of the cosmic infrared background}\label{sec:cirb}

Our knowledge of the Cosmic InfraRed Background (CIRB) at both near and far infrared 
wavelengths have been drastically increased by observations with COBE (\cite{hau00},
\cite{lag99}, \cite{wri04}) and the Japanese IRTS mission \citep{matsu05}.
In the near infrared, the observed 2~$\mu$m brightness of CIRB is more than a factor of
two larger than that estimated by ultra-deep galaxy counts based on large 
ground-based telescopes (\cite{tota01}, \authorcite{matsu05}), suggesting the
existence of a contribution from a new population, such as the first stars (\citet{kas05}
and references therein).
From the IRTS results  \authorcite{matsu05} found a signiture of a
break at 1-2~$\mu$m in the SED of the CIRB, and the existence of a bump in the power spectrum of 
the spatial fluctuation of the CIRB. 
The IRC onboard AKARI has spectroscopic capability in the near infrared 
 to $\geq 2\,\mu$m.  Moreover for this study, the IRC 
is very powerful because of its higher spatial resolution than those of COBE and IRTS as well as the high point-source 
sensitivity which will enable us to subtract any foreground point-like sources (stars and galaxies). Hence we will be able to investigate the small scale spatial fluctuations of the residual background. 
With IRTS, an appreciable spatial correlation at the 1-2~degree scale has been reported, therefore at least a
 survey scale of 2-4~degrees is required to obtain good statistical
samples of such correlation signals. Supplemental data at various ecliptic 
latitudes are also necessary to subtract the contribution of the zodiacal light.

\begin{figure}[htbp]
  \begin{center}
    \FigureFile(160mm,55mm){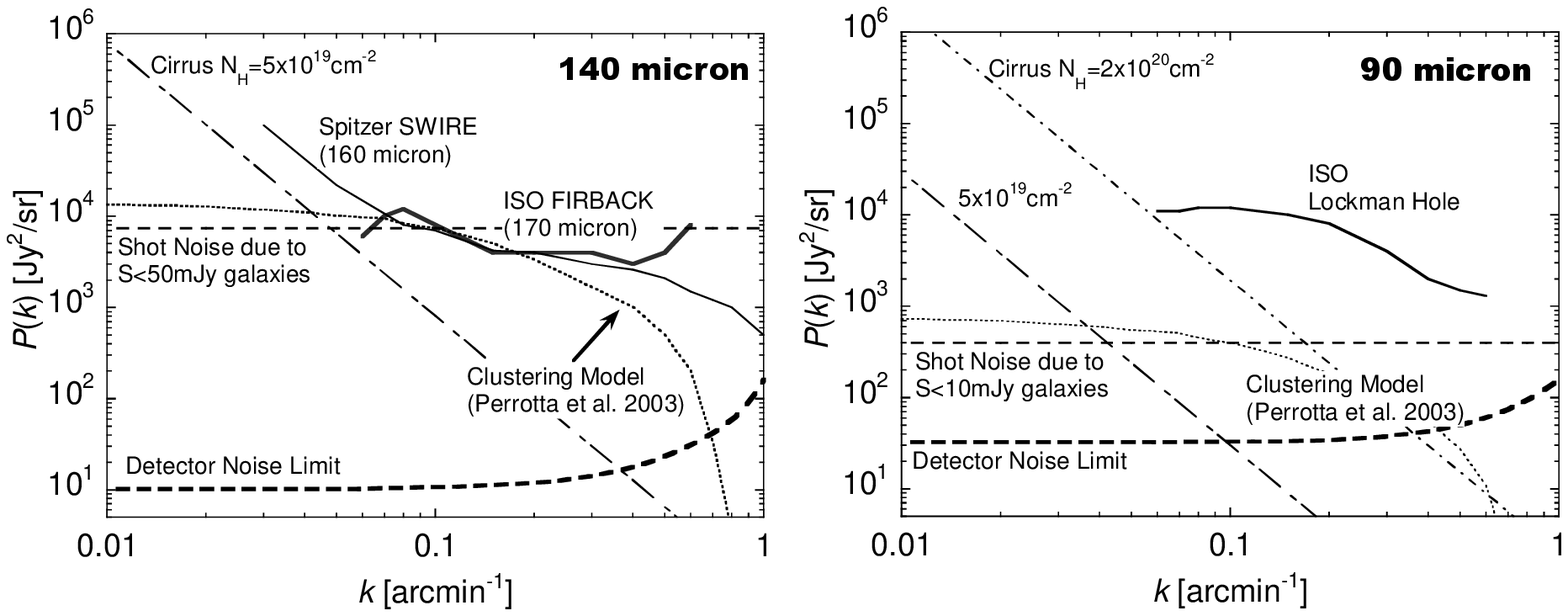}
  \end{center}
  \caption{Angular power spectrum of the CIRB fluctuation in the far-infrared. Results of the
 ISO/FIRBACK 170$\mu$m survey in ELAIS N2 \citep{pula01}, ISO 90~$\mu$m survey in Lockman Hole 
(\cite{maruma00}): $N_{\rm HI} < 10^{20}$~cm$^{2}$, and recent preliminary analysis of Spitzer/SWIRE
 160~$\mu$m survey \citep{gros05}, are also shown. Shot noise (thin dashed lines) is 
estimated for randomly distributed background galaxies fainter than 10 and 50~mJy at 90 and 
160~$\mu$m, respectively.
The data at 160(170)~$\mu$m show better agreement with the galaxy clustering model 
(dotted, \cite{perr03}) than the shot noise case. For comparison, the cirrus power 
spectra for $N_{\rm HI}=5\times10^{19}$~cm$^{2}$ (in the low-cirrus region 
near SEP) and $2\times10^{20}$~cm$^{2}$ are also shown by dash-dotted lines. The AKARI sensitivities 
(thick dashed lines) calculated by assuming a white noise spectrum correctecd for the spectra of the point-spread
function are far below the expected background fluctuation power. 
 }
\label{angpwr}
\end{figure}

 In the far infrared, not only the brightness of the CIRB but also its spatial 
fluctuations have been studied by observations with ISO (\cite{lag00}, \cite{maruma00}).
However the source confusion limit is more serious 
than in the near-to-mid infrared due to the larger diffraction-limited beam size, inhibiting the
detection of high redshift sources.  On the other hand, the fluctuation power of 
the surface brightness of the sky, after excluding the bright point sources, contains
information on the sources located at relatively high redshift. Hence, the fluctuation study 
is a very powerful tool  to understand the nature of the sources responsible for the CIRB. 
Figure~\ref{angpwr} shows the angular power spectra of the
surface brightness fluctuations in the far infrared, observed with {\it ISO} and 
{\it Spitzer}. These power spectra are calculated from the images where $S \geq 50$ mJy 
sources are removed, and are consistent with the shot (Poisson) noise due to $S < 50$ mJy sources.
Moreover, the fluctuation power is consistent with the view of strong 
evolution in the galaxy counts at 50-200 mJy, which can be interpreted as the emergence 
of a huge number of Ultra-luminous or Hyper-luminous infrared galaxies at $z\sim1$ 
(\cite{pea01}, \cite{takeu01}). 

Not only the Poisson noise arising from the discrete nature of the galaxies, but also the 
angular correlation (clustering) of the galaxies are expected to contribute to the power of the
surface brightness fluctuations (\cite{peeb80}). The observed fluctuation power, however,
shows no evidence for any contribution from galaxy clustering. Thus, the major scientific 
goal of the fluctuation studies with AKARI and {\it Spitzer} is to confirm the
existence of the fluctuation power due to clustering, and to estimate the redshift
of the galaxies responsible for the clustering. For this purpose one should perhaps examine
the cross-correlation of multi-color images in order to isolate the fluctuation power 
due to sources over some redshift range (say, $z \sim 1$ or $z \sim 2$). The signal-to-noise
ratio of the images should be high enough to obtain the fluctuation power due to $z \sim 2$
ULIRGs ($S\sim 10$ mJy). Clearly the AKARI all-sky survey is not adequate since the flux
limit is higher than the source confusion limit, and thus the slow-scan mode observation
is mandatory (see Table~\ref{tab:bandfl}, \ref{tab:conf}). Figure~\ref{angpwr} also shows
the expected contribution of the detector noise in the slow-scan observation mode. 
Regarding the required area for the
survey, one should also take into account of the field-to-field variation of the clustering
signal ({\it i.e.} cosmic variance). As we have shown in the previous subsection, 
a survey area of order 10~deg$^2$ is required to investigate the clustering due to $z\sim1$ galaxies.
The low-frequency noise due to the Galactic infrared cirrus (cirrus noise) should be also 
minimized and is discussed in subsection~\ref{sec:cirrus}. Moreover, the survey area should 
be contiguous, in order to confirm the existence of the cirrus noise at a few degrees scale, and
to judge whether the cirrus noise will not a problem at higher spatial frequencies.

The survey area and depth requirements to obtain the desired scientific objectives described
in this subsection are summarized as follows:
\begin{itemize}
\item survey depth: for the study of the cosmic near-infrared background fluctuations, the survey 
  area should be a contiguous, 2-4~degrees wide, circular or square shape. The 
  sensitivity requirement is easily achieved with the exposure time of a single pointing
  observation. Good stability of the detector array is a key issue to obtain a reliable
  signal of the background fluctuations at low spatial frequencies.
\item survey area: for the study of the cosmic far-infrared background fluctuations, the 
  signal-to-noise ratio of the images should be high enough to obtain the 
  fluctuation power due to $z \sim 2$ ULIRGs ($S\sim 10$ mJy at 90-140~$\mu$m),
  and hence requires the slow-scan observation mode. a survey area of order 10~deg$^2$
  is required to investigate the clustering due to $z\sim1$ galaxies.   
\end{itemize}

\subsection{Statistical study of buried AGN}\label{sec:agns}

Not only the cosmic star formation history but also the birth and evolution
of Super-Massive Black Holes (SMBHs) are major science goals of modern
astronomy.  Moreover, the birth and evolution of the SMBHs is closely related 
to the evolution of the host galaxies, as indicated by the tight correlation between the
mass of SMBHs and the velocity dispersion of the bulge component of host galaxies 
({\it e.g.}  \cite{mefe01}) in the local universe. Furthermore, results from 
recent hard X-ray surveys have revealed that most of the accretion onto the SMBHs 
is obscured by dust \citep{ued03} prohibiting a diagnostic study by 
optical lines (``buried AGN"). In such buried AGNs most of the AGN luminosity is 
absorbed by dust and re-radiated in the mid to far-infrared. Thus in order to unveil 
the nature of such buried AGN, infrared observations with AKARI will play 
a key role. Moreover, to discriminate between the hot dust emission powered by AGNs 
and the warm dust emission powered by star formation, multi-color data at optical to 
far-infrared wavelengths is particularly useful. 
This will also enable us to understand the AGN contribution to the CIRB, especially at 
far-infrared wavelengths.

As described in section~\ref{sec:dusty} the proposed NEP survey will provide a
huge number of ULIRGs or Hyper-luminous infrared galaxies (HLIRGs) out to $z=3$, and these 
samples will provide a unique database to investigate the contribution of the buried 
AGN as the energy source for the ULIRGs. One should note that such studies
will be more effective when observing time is allocated to
XMM/Newton hard X-ray imaging of the NEP AKARI deep field. 

Another diagnostic tool to investigate the energy sources of ULIRGs are
the dust features in the thermal infrared. Since the energy source for the 
buried AGN is, by definition, concentrated at the center of the dust cloud,
its presence is traced by the carbon dust absorption feature at 3.4~$\mu$m and
the 10-20~$\mu$m silicate dust features. So far 6-11~$\mu$m spectra of many
ULIRGs with ISO \citep{rig99} and ground-based $L$-band spectra
\citep{imad00} have been obtained. However, the limited wavelength coverage
made it very difficult to determine whether the spectra were dominated by strong
7.7~$\mu$m polycyclic aromatic hydrocarbon (PAH) emission features 
(starburst-powered) or a strong 9.7~$\mu$m absorption
feature of silicate dust implying the presence of a buried AGN.  

The AKARI/IRC incorporates slitless spectroscopic functions at 2-26~$\mu$m with a
spectral resolving power of 30-90, suitable for the study of relatively
broad spectral features. Therefore, we will perform an unbiased spectroscopic 
survey by using the slitless spectroscopic mode in order to provide numerous
SED samples of various galaxy types including the serendipitous detection of buried 
AGN candidates. 
Moreover, it is noted that the studies with 3.3~$\mu$m PAH feature of local starburst
galaxies as well as with the 3.4~$\mu$m feature of local AGNs are impossible 
with {\it Spitzer}/IRS, due to the lack of wavelength coverage below 5~$\mu$m.

The survey area and depth requirements to obtain the desired scientific objective described
in this subsection are summarized as follows:
\begin{itemize}
\item survey depth and area requirements for the imaging surveys are already stated in the
 summary of section~\ref{sec:dusty}. The buried AGN population is an important
 subset of the entire galaxy sample. Detailed discussion will be given in a separate
 paper by Pearson et al. (2006).
 \item survey depth and area requirements for the spectroscopic survey: good sensitivity
 for the mid-infrared dust features is a key issue in order to generate a sample of unbiased mid-infrared SEDs of galaxies in sufficient 
 numbers . This will be also discussed
 in a separate paper.
\end{itemize}

\section{Optimization of the Location of the Survey Fields}\label{sec:optimize}

In this section, we describe the best location of the survey area for the extragalactic
deep survey with AKARI.
In short, we first examined the visibility constraint of the AKARI spacecraft which
leads us to conclude that the ecliptic pole regions are the almost unique choice, and then 
checked if the survey will suffer from the confusion noise due to the Galactic 
infrared cirrus, and also investigated the density of bright sources (mostly stars 
in the Galaxy) which may cause serious saturation of the near-infrared detectors. 

\begin{figure}[ht]
  \begin{center}
    \FigureFile(110mm,110mm){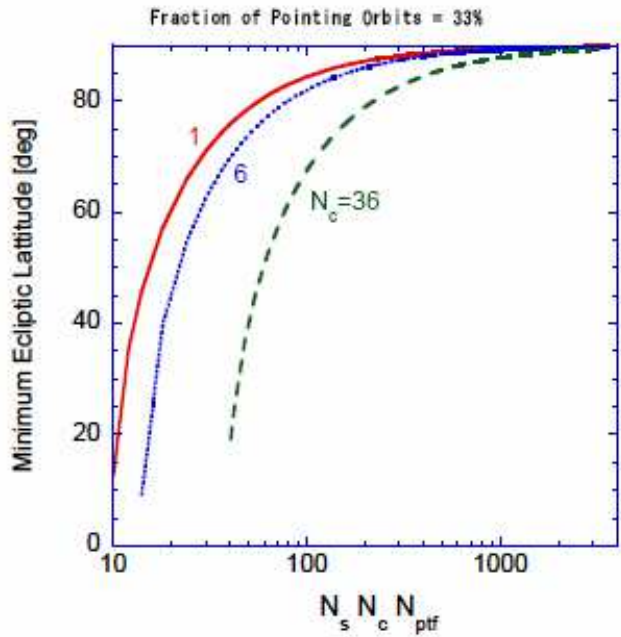}
  \end{center}
  \caption{The minimum value of ecliptic latitude ($\beta$)  as a function of 
the number of IRC pointing observations. The figure shows the minimum latitude 
at which surveys of $N_{\rm c}$=1, 6, 36  number of Field of Views (FOV) in the cross 
scan direction can be made. The FOV of the IRC is approximately 10$\times$10~arcmin$^2$, so 6 
FOVs  in the cross scan direction ($N_{\rm c}$=1, 6) corresponds to 1 degree.  
The number of pointing observations in the in-scan direction are given by $N_{\rm s}$
and the number per FOV (i.e. the depth) is given by $N_{\rm ptf}$. }\label{point}
\end{figure}

\subsection{Visibility constraint of AKARI}

According to the attitude control scenario described in section~\ref{sec:AKARI}, 
we checked the sky visibility constraint for the deep surveys which require
a significantly large number of pointed observations. 
For simplicity we assume a rectangle survey area of uniform depth at an ecliptic 
latitude $\beta$, satisfying the following equation :

\begin{equation}
N_{\rm c}N_{\rm s}N_{\rm ptf} \leq {10N_{\rm c} + 120 \over{ \Delta \lambda_{0}
 \cos \beta}},  \label{minbeta}
\end{equation}
\noindent
where $N_{\rm c}$, $N_{\rm s}$, and $N_{\rm ptf}$ are the number of FOV in the cross-scan 
direction, number of FOV in the scan direction, and the number of pointed observations per 
field-of-view respectively. $\Delta \lambda_{0} = 4.1~$arcmin is the longitudinal
separation between successive orbits at the ecliptic plane. This equation gives 
the minimum $\beta$ at which we could perform a survey with a total number of pointed observations
$N_{\rm c}N_{\rm s}N_{\rm ptf}$, and is shown in Figure~\ref{point}.

For example, 1~deg$^{2}$ coverage with 30 pointings per FOV requires 
$N_{\rm c}N_{\rm s}N_{\rm ptf} \sim 1000$.
Clearly such a required number of pointing opportunities 
can only be allowed in the regions very close to ecliptic poles ($\beta \geq 89$~deg) unless
the survey area is a long strip or a thin donut. Such geometries do
not satisfy the requirement from the scientific objectives described in 
section~\ref{sec:objectives}. Hence the deep survey field must be chosen at the
ecliptic polar regions.

\subsection{Confusion due to the infrared cirrus emission}\label{sec:cirrus}

An important sensitivity limitation arises due to the spatial fluctuation of the IR 
cirrus. \citet{hebe90} have assumed that the fluctuation power is proportional
to $ \bar{B}_\nu^{3}$ for all wavelengths, where $\bar{B}_\nu$ is mean brightness of the IR cirrus.
They have expressed the cirrus confusion noise in the following equation :

\begin{equation}
{\sigma(\lambda) \over{\rm 1\,mJy}} = 0.73 ({\lambda \over{100\,\mu \rm m}})^{2.5} 
({D \over{\rm 0.7m}})^{-2.5} ({\bar{B}_\nu \over{\rm 1\,MJy sr^{-1}}})^{1.5} \hskip 12pt .
  \label{heloubeichman}
\end{equation}

Thus the cirrus confusion noise will depend not only on the variation of the surface 
brightness of the background structure, but also on the resolution of the telescope ($\lambda / D$),
resulting in less noise at shorter wavelengths.  \citet{kis01} have 
analyzed 40 sky regions with the ISOPHOT instrument on ISO concluding that the 
cirrus noise was consistent to within a factor of 2 according to the formulation of \authorcite{hebe90}.

\begin{figure}[htbp]
  \begin{center}
    \FigureFile(160mm,80mm){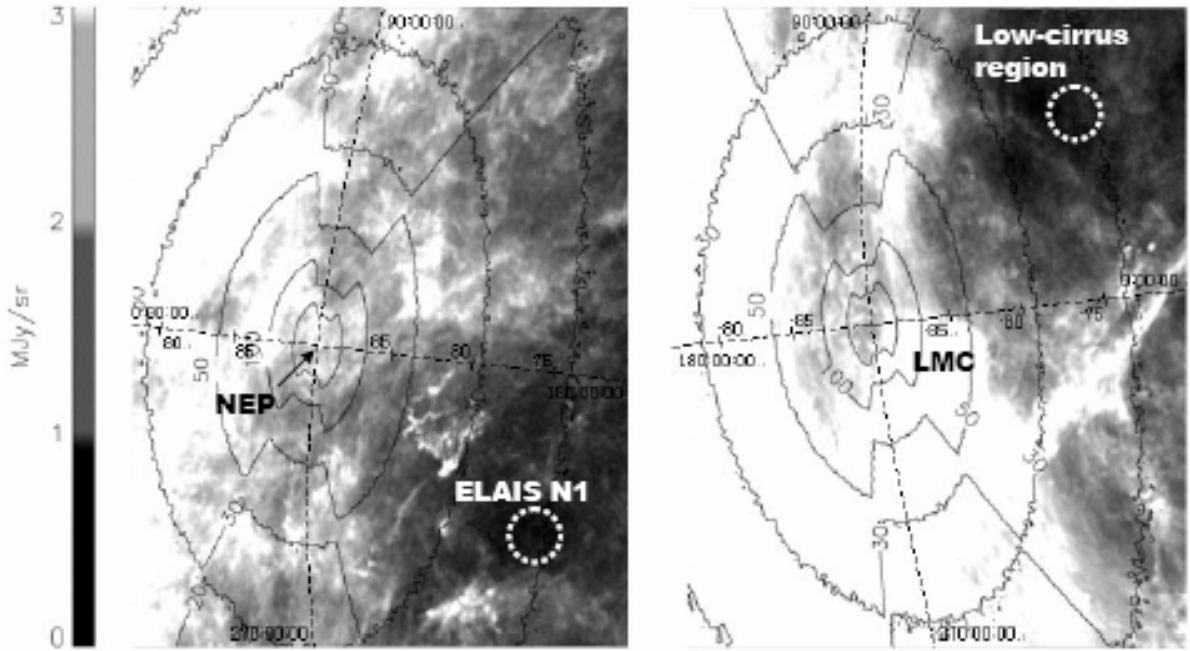}
  \end{center}
  \caption{IRAS 100$\mu$m map near the NEP (left) and SEP (right), overlayed on the 
visibility contours. Darkest areas are approximately 0.3~MJy/sr, and brightest areas
correspond to 3~MJy/sr. ELAIS N1(a {\it Spitzer}/SWIRE field) and the low-cirrus regions 
from \citet{sch98} are indicated. }
\label{nepseplc}
\end{figure}

Figure~\ref{nepseplc} shows the IRAS 100~$\mu$m map (\cite{sch98}) near 
the ecliptic poles. Unfortunately, as shown in the figure, the ecliptic poles
are not optimal areas for the local minimum of the infrared cirrus: 2-3~MJy sr$^{-1}$.
Applying equation~\ref{heloubeichman} to the AKARI FIS/IRC 
wavebands and assuming a range of mean brightness for the infrared cirrus at
100~$\mu$m ($\bar{B}_{100}$)  with the 
infrared cirrus SED of \citet{dwek97}, we obtain the results for the cirrus 
confusion noise in Table~\ref{tab:conf}. Unless a relatively bright cirrus 
region ($\bar{B}_{100}\geq 5\, \rm MJy\,sr^{-1}$) is observed, the cirrus 
confusion is not problematic for IRC bands. It is noteworthy that, based on recent 
{\it Spitzer} observations at 24~$\mu$m in a bright cirrus region in Draco ($N_{\rm HI}=4-14
\times 10^{20}\,\rm cm^{-2}$), \citet{dol04} reported a rather weak effect 
leading to a completeness degradation of only 15\%.

However, the cirrus noise toward the ecliptic poles is appreciable for the FIS bands 
(especially LW channels). Hence, we should carefully 
choose any deep field for FIS imaging, especially for the studies of CIRB fluctuations. One should
also take into account any angular frequency dependence of the fluctuation power spectrum
of the cirrus : $P(k) \propto k^{-3}$ (\cite{Gau92}, \cite{kis01}, \cite{jeon05}), where
$k$ is the angular frequency. Fluctuations at large angular scales (order of a degree)
are likely to be dominated by cirrus noise, even though the point source detection
({\it i.e.} at high angular frequency) is not affected. Hence a low-cirrus region is 
required, especially for the CIRB study in the far-infrared. The 
optimum area for such far-infrared deep surveys is the low cirrus region in the South
or in the ELAIS N1 field ($\bar{B}_{100}\leq 0.5\, \rm MJy\,sr^{-1}$, corresponding to a hydrogen 
nuclei column density of $5\times10^{19}$~cm$^{-2}$) as shown in 
Figure~\ref{nepseplc}. Figure~\ref{angpwr} shows the expected cirrus power spectra toward
the low cirrus region near the SEP, compared with the ``shot noise" due to the unresolved 
galaxies. The advantage of the low cirrus region near the SEP can be clearly seen. 

\begin{table}
  \caption{Estimated confusion limits in AKARI wavebands. Cirrus noise is calculated assuming
   the model of Helou \& Beichman (1990)}\label{tab:conf}
  \begin{center}
    \begin{tabular}{lllllll}
	\hline \hline
Band     & $\lambda_{\rm c}$  & \multicolumn{2}{l}{Galaxy confusion (5$\sigma$ $\mu$Jy)} & 
   \multicolumn{3}{l}{ Cirrus Confusion  (5$\sigma$ $\mu$Jy) for $\bar{B}_{100}$ }   \\
(IRC)    &  ($\mu$m)     & 40~beam/source & 20~beam/source &  0.5~MJy/sr  &  3~MJy/sr & 10~MJy/sr \\ \hline
N2       &    2.43       &  $<$1         &   $<$1        &    $<$1      &   $<$1    &   $<$1    \\ 
N3       &    3.16       &  $<$1         &   $<$1        &    $<$1      &   $<$1    &   $<$1    \\ 
N4       &    4.14       &  $<$1         &   $<$1        &    $<$1      &   $<$1    &   $<$1    \\ 
S7       &    7.3        &  1            &   $<$1        &    $<$1      &   $<$1    &   $<$1    \\ 
S9W	 &     9.1        &  3            &   $<$1        &    $<$1      &   $<$1    &   1.1     \\ 
S11      &    10.7       &  7.9          &   2           &    $<$1      &   $<$1    &   1.6     \\   
L15      &    15.7       &  50           &   16          &    $<$1      &   1.1     &   6.9     \\ 
L18W	 &     18.3       &  134          &    66         &    $<$1      &   1.6     &   9.5     \\ 
L24      &    23.0       &  257          &   134         &    $<$1      &   3.1     &   19.     \\ \hline
Band     & $\lambda_{\rm c}$  & \multicolumn{2}{l}{Galaxy confusion (5$\sigma$ mJy)} & 
   \multicolumn{3}{l}{ Cirrus Confusion  (5$\sigma$ mJy) for $\bar{B}_{100}$ }   \\
(FIS)    &  ($\mu$m)     &               &               &  0.5~MJy/sr  &  3~MJy/sr & 10~MJy/sr \\ \hline
N60      &      65       & \multicolumn{2}{l}{\hskip 2cm 3.0} & 0.06    &   0.91    &   5.5     \\ 
WIDE-S   &      90       & \multicolumn{2}{l}{\hskip 2cm 7.0} & 0.33    &   4.8     &   29.     \\
WIDE-L   &     140       & \multicolumn{2}{l}{\hskip 2cm 45} &  9.2     &   135.    &   823.    \\
N160     &     160       & \multicolumn{2}{l}{\hskip 2cm 50} &  12.     &   173.    &  1050.    \\ \hline
    \end{tabular}
  \end{center}
\end{table}

\subsection{Bright Stars} 

\begin{figure}[htbp]
  \begin{center}
    \FigureFile(150mm,90mm){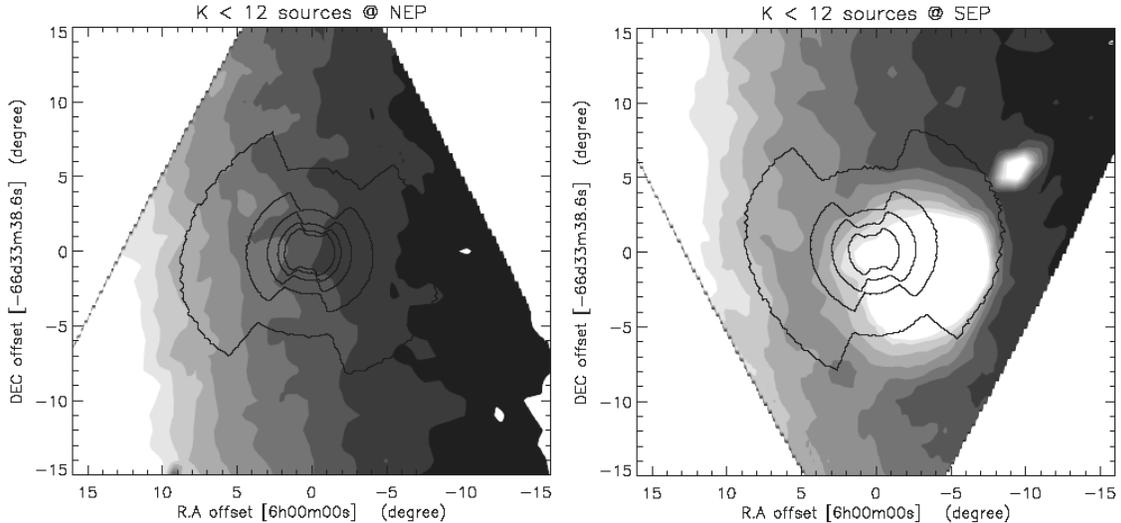}
  \end{center}
  \caption{Bright K$<$12 2MASS source density around NEP region(left) and SEP region(right). 
AKARI visibility contours (solid lines) are shown for 250, 200, 
150, 100, 50 (in phase 1 plus 2) away from the pole. Gradient corresponds to 
130(darkest), 160, 190, 220, 260, 290, 320, 350, 380, 410 stars per square 
degree.}\label{nepsep2m}
\end{figure}

Bright stars in our Galaxy as well as the Magellanic clouds should be avoided for extragalactic deep surveys. 
Not only the saturated pixel due to the bright stars
but also the adjacent pixels may not be useful for deep imaging because such pixels are
affected by an increased photon noise. $K\leq12$~mag corresponds
to the saturation limit of the N2 band imaging of IRC. However, since the FOV of the IRC is as wide as
$10'\times10'$, it is difficult to find fields without any bright stars in the ecliptic
polar regions. In Figure~\ref{nepsep2m}  the density distributions of bright 2MASS stars
 ($K\leq12$~mag) are shown with the AKARI visibility contours.  At the NEP approximately 5 
such stars exist per FOV of the IRC. Below $K\simeq 15$~mag extragalactic sources 
begin to dominate the source counts, and the photon noise due to the stars is not
appreciable. By using the star count model of the Galaxy \citep{naka00}, we estimated approximately
on average 50 stars per FOV down to $K\sim15$~mag.
When considering the current best estimate for the in-flight point spread function of the IRC-NIR
channel (approximately 10 pixels per source in the 70\% energy circle),
this means approximately only 0.3\% of  pixels will suffer from saturation or the 
increased photon noise due to stars. In the case of the SEP, as shown in Figure~\ref{nepsep2m} the surface density
of the Galactic stars is about twice larger than that at NEP since it is close to the Large Magellanic
Cloud. In conclusion, the surface density of the Galactic stars is non-negligible but still within acceptable limits
to perform the extragalactic surveys. Thus for the near and mid infrared deep surveys with AKARI, 
we select the field near NEP simply because the star density is a factor of two smaller at the NEP, 
and moreover, we cannot allocate so many pointed observations for extragalactic
surveys near SEP because of the existence of another large-area survey with AKARI: the LMC survey.

The bright stars will give serious damages on the optical pre-survey images. Hence the field position 
was chosen to avoid saturation of the CCDs of the Suprime-cam instrument by bright ($V \leq 10$~mag) stars
(see section~\ref{sec:ground}). 

\section{Observation Plan}\label{sec:plan} 

\subsection{Proposed Area for the deep potinting surveys}

From the above discussion, regarding the IRC deep survey field, the NEP is the best candidate because 
approximately 1000 pointing observations can only be realized within a few degrees from NEP
(see Figure~\ref{point}), and there are no other serious limitations. There is an option to move
 ``NEP-Wide" survey field to ELAIS N1 field, one of the {\it Spitzer}/SWIRE fields. However, 
the ecliptic latitude of the ELAIS N1 is not high enough to be observable in all the year, 
and thus it is hard to generate a circular or nearly rectangular map which is one of the key 
requirements for the study of the cosmic near-infrared background fluctuations 
(see section~\ref{sec:cirb}). In contrast, the SEP is 
not so good for the IRC extragalactic deep survey due to its proximity to the LMC, and moreover,
there exists a large-area survey program toward the LMC which makes it difficult to allocate a
large number of pointed observations for any extragalactic survey. Thus, for the near and 
mid infrared deep extragalactic surveys we conclude the NEP is the best target field.

Regarding the far infrared deep surveys, the NEP is acceptable since
the cirrus confusion limit is approximately the same or below the galaxy confusion limit
considering a mean cirrus brightness (at IRAS 100~$\mu$m) of $\bar{B}_{100}= 1-2$~MJy/sr and
$\bar{B}^{3}$ dependence of the fluctuation power spectrum (\cite{Gau92}, see 
Table~\ref{tab:conf}). However, an extremely-low cirrus region is preferable
since the major scientific goal,  measurement of the CIRB fluctuations over degree scales, will be
hampered by the $k^{-3}$ dependence of the infrared cirrus emission, where $k$ is angular 
frequency. As shown in Figure~\ref{nepseplc} alterantive candidate fields are the ELAIS N1 field near 
the NEP and the low-cirrus region ner the SEP at $\beta=-74$~deg, where $\bar{B}_{100}$ is as 
low as 0.2~MJy/sr \citep{sch98}. The advantage of the ELAIS N1 field near the NEP is
that there are numerous multiwavelength data sets already
available including the {\it Spitzer} data. However, pointed observations for the NEP survey (with the IRC)
severely conflicts with those for the ELAIS N1 fields, and a large-area survey (more than 10~deg$^2$)
is not realistic. On the other hand, pointing opportunities for the SEP low-cirrus region
do not conflict with those for the LMC survey ({\it i.e.} the locations are different in 
ecliptic longitude), and thus we conclude that the SEP low-cirrus region is the best 
target field for the far infrared deep slow-scan surveys.  In Table~\ref{tab:epsvy} we
summarize the overview of the currently planned deep surveys.

\subsection{The NEP survey}\label{subsec:nep} 

In this subsection we describe the details of the current observation plan
for the NEP survey with AKARI in its pointing attitude mode. In order to 
satisfy the scientific requirements (section~\ref{sec:objectives}), the survey
area should be observed to a suitable depth for the detection of
statistically meaningful samples of extragalactic sources ($\gg$1000)
out to $z$=4 in the near infrared, and to $z$=3 in the mid infrared. Moreover,
the survey area should be large enough so that the obtained sample does not
suffer from serious uncertainty due to cosmic variance, and the geometry 
of the area should be a circle or a square, {\it i.e.} a strip or ring 
should be avoided. 

Currently the following two blank-field surveys are planned:
\begin{itemize}
\item {\bf NEP-Deep:} deep (28 pointing observations per field-of-view)
  survey over approximately 0.5~deg$^{2}$
\item {\bf NEP-Wide:} wide and shallow (2 pointing per FOV) survey 
  of approximately 6.2~deg$^{2}$. The survey region is a circular area
  surrounding the ``NEP-Deep" field.
\end{itemize}

\begin{figure}[htbp]
  \begin{center}
    \FigureFile(160mm,115mm){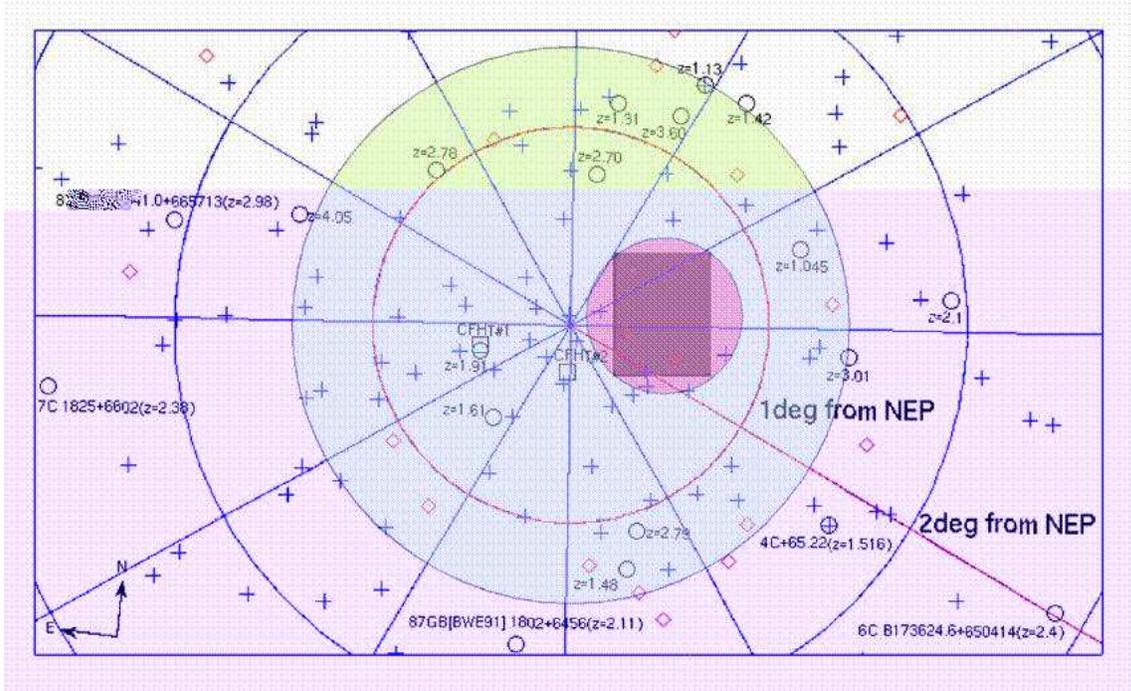}
  \end{center}
  \caption{Survey area planned for the NEP survey in pointing mode, 
	consisting of a circular area (``NEP-Deep", smaller one surrounding the rectangle) with 28 pointing per
	FOV and a larger circular area (``NEP-Wide", centered at NEP) with 2 pointings per FOV.
	Also shown are the optical deep survey field with Subaru/suprime-cam (rectangle area) as well as
	the location of ROSAT X-ray clusters (red diamonds, \cite{gio03}), ROSAT galaxies,
	AGN, planetary nebulae (blue crosses), and high-$z$ 
	radio sources (circles, \cite{bri97}). 
 }\label{neparea}
\end{figure}

For the former, the one-filter AOT will be used (8-10 pointing observations per filter
and per field-of-view) while for the latter, the three-filter AOT will be used.
Therefore data will be obtained in all 9 IRC wavebands.
The locations of the survey fields is shown in Figure~\ref{neparea}. The total number of
pointed observations amounts to 954. During phase-1 and the first six months of
phase-2,  approximately 2.6 pointing opportunities per day using mainly the orbits
passing the SAA will be allocated for the survey.
Pointing directions are primarily determined by the requirements from
the IRC observations: the FIS will be operated in the FTS mode (probably in
low spectral resolution mode of approximately 1~cm$^{-1}$) in parallel, in
order to obtain the 50-180~$\mu$m SED of the diffuse emission.  Although the FTS
mode is not sensitive enough to obtain the spectrum of each extragalactic source, these FTS observations
toward the NEP are useful for the study of the CIRB as well as Galactic interstellar matter, by using
several hundreds of pointing opportunities in this parallel mode.

\subsubsection{The Deep survey ``NEP-Deep"}

The survey area of ``NEP-Deep" is a circular area whose center is slightly
offset from the NEP, in order to cover the deep pre-survey field which has
been already surveyed at optical wavelengths with Subaru/Suprime-cam 
(see section~\ref{sec:ground}). The survey field
is observable for many orbits during the mission, when the offset
control is applied. This will be important if we are to cover the field with
all IRC channels efficiently. This reasoning is explained as follows: first, as
shown in Figure~\ref{fov} the FOV of the MIR-L channel is offset from those of 
the NIR and MIR-S channels by approximately 20 arcmin, thus one pointing 
observation provides a NIR/MIR-S image and a MIR-L image separately
on the sky. Secondly, since the FOV configuration rotates 180~deg in approximately
6 months, another pointing observation after 6 months enables us to obtain
the corresponding images in the complementary channel.
Then two $10'\times10'$ fields separated by approximately 20 arcmin have 
been covered with all IRC channels.
Therefore, allocation of the same number of pointed observations in both phase-1
and phase-2 (first 6 months) is a mandatory requirement for the ``NEP-Deep" survey.
Since the 0.5~deg$^2$ corresponds to 18 independent
FOV, 28 pointing per a FOV results in 504 pointed observations in total with just
one half of them being executed in Phase-1.
 Table~\ref{tab:nepdepth} summarizes the expected flux limits for the NEP
survey. 

\subsubsection{The Shallow-Wide survey ``NEP-Wide"}

The survey region of ``NEP-Wide" is selected to surround the ``NEP-Deep" field but 
is a circular area whose center coincides with the NEP. Unlike the 
``NEP-Deep" the offset control is not actively used, but the pointing directions
are regularly placed ({\it i.e.} FOVs are primary ordered in the in-scan direction, 
then the sky coverage is expanded in the cross-scan direction as the scan path 
rotates) so that the every portion of the survey field is covered with 
a uniform depth of 2 pointing per FOV in three filter mode (AOT-03). In this 
manner the survey field will be covered by both the NIR/MIR-S and the 
MIR-L  over a year (phase-1 plus the first six months of phase-2). 
Currently we plan to allocate 450 pointed
observations in total, resulting in 225 independent FOVs corresponding to 
approximately 6.2~deg$^2$.

The driving force of the ``NEP-Wide" survey is to perform a survey of as large area as 
possible especially in N2 band of the IRC in the search for the large-scale fluctuations
of the cosmic near infrared background (section~\ref{sec:cirb}). For the mid
infrared channels, the main purpose 
of ``NEP-Wide" is to overcome effects due to the cosmic variance at $z$=0.5-1: as 
shown in Figure~\ref{comvol} the survey volume of ``NEP-Deep" is clearly too small.
In addition,  it is also noteworthy that more than 1000 ultra or hyper-luminous infrared galaxies ($L_{\rm IR} \geq 10^{12}\, L_{\odot}$) per square degree
 will be detected by this 
shallow-wide survey (250$\mu$Jy at 24$\mu$m, \cite{pea05})  at $z=2-3$ .

\begin{table}
  \caption{IRC flux Limits expected for the NEP survey}\label{tab:nepdepth}
  \begin{center}
    \begin{tabular}{llll}
	\hline \hline
Band     &  $\lambda_{\rm c}$ & NEP-Deep\footnotemark[\#] & NEP-Wide\footnotemark[\$] \\
(IRC)    &  ($\mu$m)          & (5$\sigma$, $\mu$Jy) & (5$\sigma$, $\mu$Jy) \\ \hline
N2       &    2.43            &  2.8                 &    12                \\ 
N3       &    3.16            &  1.2                 &     5.5              \\ 
N4       &    4.14            &  2.3                 &    11                \\ 
S7       &    7.3             &  10                  &    49                \\ 
S9W	  &    9.1             &   9                  &    39                \\ 
S11      &    10.7            &  12                  &    56                \\ 
L15      &    15.7            &  22                  &   100                \\ 
L18W	  &     18.3           &  31                  &   130                \\ 
L24      &    23.0            &  57                  &   270                \\ \hline
Band     &  $\lambda_{\rm c}$ & (slow-scan)\footnotemark[$*$]  & (all-sky survey) \footnotemark[$\dagger$]  \\
(FIS)    &  ($\mu$m)          & (5$\sigma$, mJy)     & (5$\sigma$, mJy)     \\ \hline
N60      &    65              &   35                 &    100               \\
WIDE-S   &    90              &    7                 &    20                \\
WIDE-L   &    140             &   4.5                &    40                \\
N160     &    160             &    9                 &    70                \\ \hline
         &                    &                      &                      \\
\multicolumn{4}{l}{\parbox{85mm}{\footnotesize \noindent
  \footnotemark[\#] AOT-00 (single filter per pointing). Net exposure times : 4000-5000~sec per filter. 
  \par \noindent
  \footnotemark[\$] AOT-03 (3 filters per pointing). Net exposure times : 200-230~sec per filter.
  \par \noindent
  \footnotemark[$*$] assumes four slow-scans with a speed of 15"sec$^{-1}$ at each sky position. 
  \par \noindent
  \footnotemark[$\dagger$]  achieved by integrating the all-sky survey data of all overlapping scans 
  toward the ``NEP-Wide" field.
 }} 
    \end{tabular}
  \end{center}  
\end{table}

\subsubsection{The spectroscopic survey}

In addition to the broad-band imaging surveys, slitless spectroscopy in all IRC channels is also planned which will provide
another unique product from the AKARI mission : an unbiased near and mid infrared
spectroscopic sample of galaxies (see section~\ref{subsec:comp}). 
IRS spectroscopic follow-up of relatively
bright ($\geq 0.75$mJy) 24~$\mu$m sources has revealed the nature and the redshift of
the sources to some extent \citep{hou05}, however, the sample is biased to relatively
bright sources.  Within a
1~deg radius circle from the NEP, an approximately 2000~arcmin$^2$ area will be 
surveyed using the spectroscopic channels (table~\ref{tab:specspec}) as part of an AKARI MP. The
details will be described in a separate paper.

\subsection{Near SEP Survey}

\begin{figure}[htbp] 
  \begin{center}
    \FigureFile(150mm,125mm){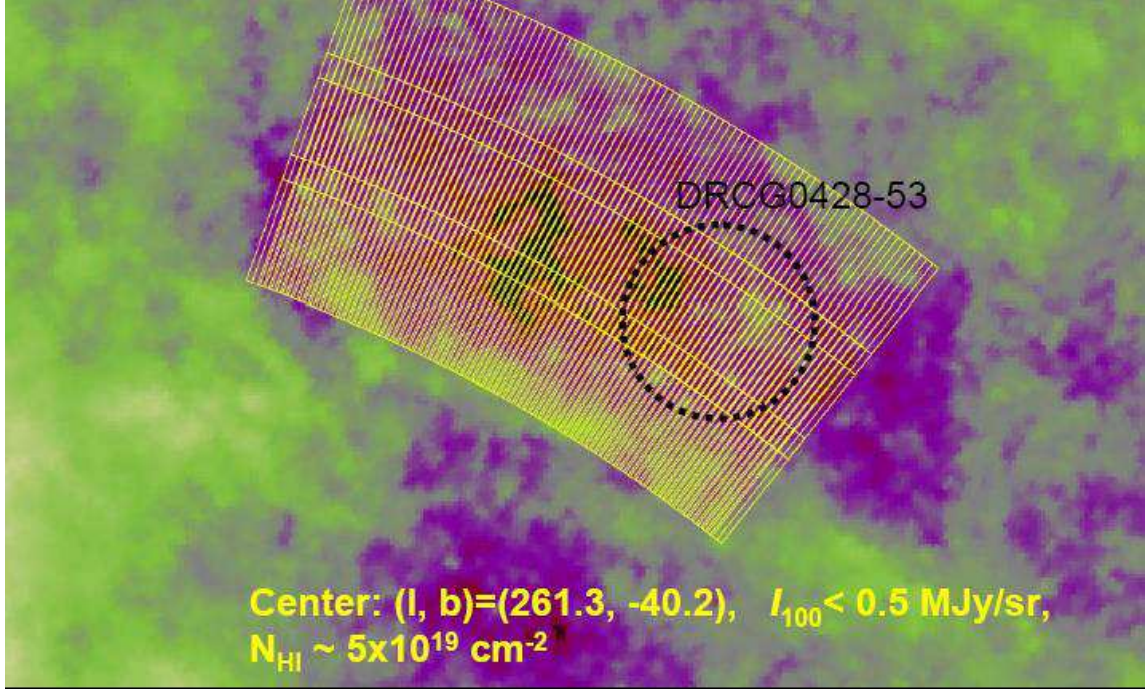}
  \end{center}
  \caption{Areal coverage planned for the low-cirrus area survey near the SEP with FIS slow-scan mode.
	The fan-shape area (approximately 15~deg$^{2}$, blue boxes) will be obtained by mosaicing
	the strip maps, each of which is obtained in one pointing opportunity.
 }\label{separea}
\end{figure}

Here we describe the details of current observing plan toward the low-cirrus region
near the SEP using the FIS slow-scan mode. The target area is shown in Figure~\ref{separea}. 
Since the major science goal of this survey is to perform a study of the CIRB fluctuations of degree
scales or larger, contiguous mapping in an area over at least 15~deg$^2$ is planned. Considering 
the visibility of the low cirrus region at $\beta \simeq 74$~deg, a daily allocation of 3-4 
pointing opportunities is mandatory to create such a contiguous map. In Figure~\ref{separea}, 
an example of the footprints of the FIS FOV are also shown as blue boxes: in this scenario, 
at each pointing observation, the FIS FOV makes a round trip (240sec forward-scan, then 
240sec backward, with $15"$/sec scan rate) along the nominal survey path, resulting an 
$8' \times 1$~deg strip map. Using 270 pointing opportunities over the 90 days when the 
low-cirrus region is visible (this corresponds to 3 pointing opportunities 
per day), 3 (along the scan) $\times$ 90 (cross-scan, with 4 arcmin steps) mosaics of 
the $8' \times 1$~deg strip maps 
corresponds to an approximately 15~deg$^2$ fan-shape area. 

\begin{figure}[htbp]
  \begin{center}
    \FigureFile(126mm,190mm){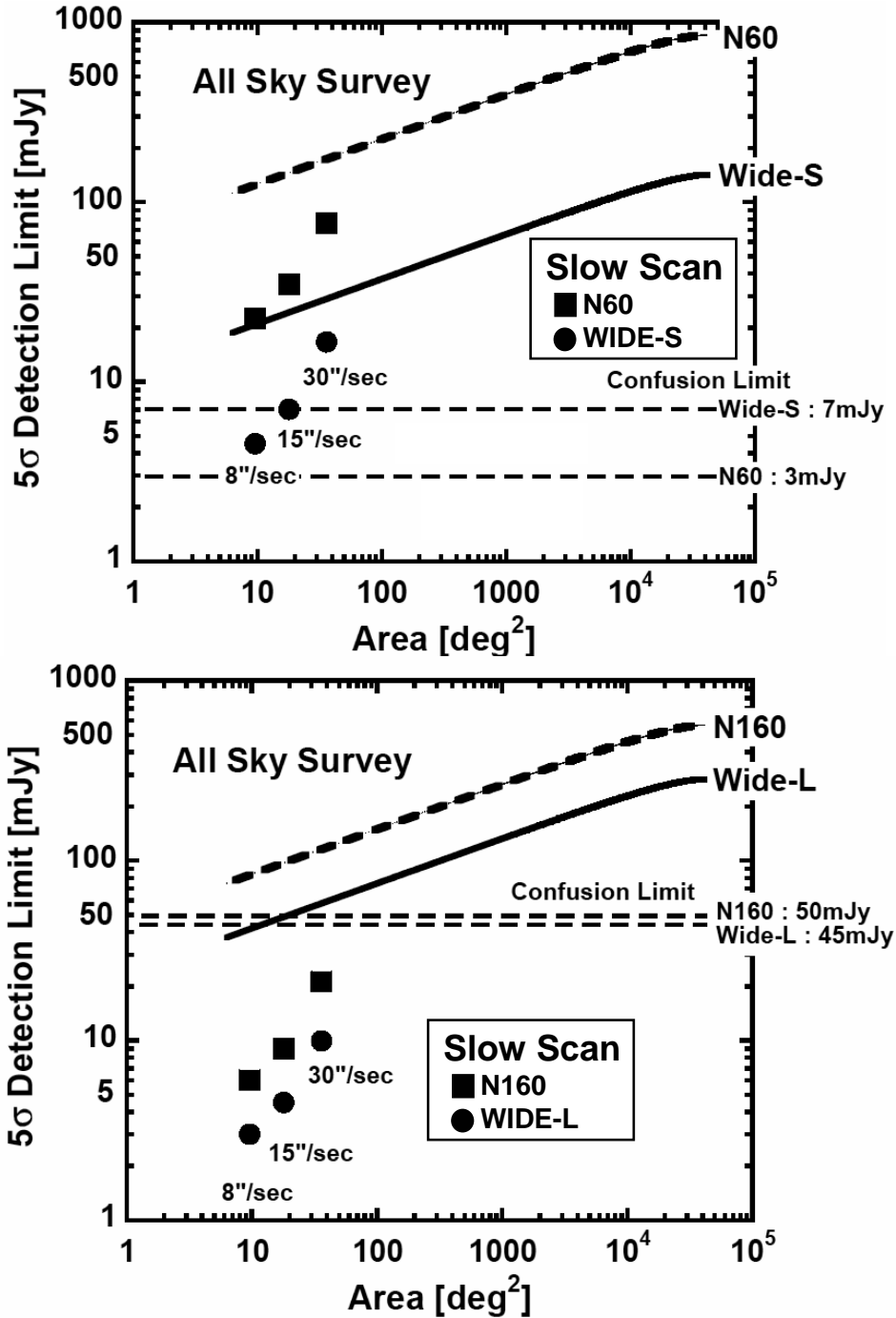}
  \end{center}
  \caption{Depth and areal coverage for various slow-scan surveys with the
	FIS bands. At each point on this plot, 270 pointing observations are assumed. 
	The resulting areal coverage varies according to the scan-speed: 8, 15, and 30 arcsec/sec. 
	For comparison, the dashed and solid curves show the all-sky survey 
	sensitivities by simply taking into account the number of overlapping scans
	at any given point on the sky. The horizontal dashed lines are the galaxy confusion limit 
	for a 69~cm telescope estimated based on the recent {\it Spitzer} results.
 }\label{depthareass}
\end{figure}

Figure~\ref{depthareass} compares the depths and areal coverage for various slow-scan rates
for a fixed number (270) of pointed observations.  The expected sensitivities of the WIDE-L and N160 bands reach
below the galaxy confusion limit even in case of the fastest scan rate (30$"$/sec), and hence
the adoption of the faster scan rate is adequate in order to expand the survey area. It is also
worth noting that the all-sky sensitivity does not reach the confusion limit even if we take into
account the large number of overlapping scans expected near the ecliptic poles.

\section{Uniqueness of the Surveys}\label{sec:uniq}

We discuss the uniqueness of the AKARI deep surveys compared to other contemporary surveys carried out 
with previous and ongoing missions.

\subsection{Infrared Space Observatory}

The {\it Infrared Space Observatory} (ISO, \cite{kes96}) launched in November 1995 
was equipped with a 60~cm cooled telescope and 
had the capability for near and mid infrared imaging: {\it ISOCAM},  a
$3'\times3'$ field of view camera with $32 \times 32 $ pixels. Most of the deep 
ISOCAM surveys utilized two broad band filters; LW2 at 6.7~$\mu$m and LW3 at 
15~$\mu$m. In particular, the ISOCAM 15~$\mu$m surveys covered a wide range in both 
sensitivity and spatial area: 0.013~deg$^{2}$ with 100~$\mu$Jy, $0.3\sim0.4$~deg$^{2}$ 
with 0.5-0.8~mJy, and 13~deg$^{2}$ with 1~mJy ({\cite{GeCe00} and references therein), resulting in the 
discovery of a strongly evolving population of galaxies below 1mJy. This evolving 
population can be interpreted as the presence of luminous, ultra-luminous and hyper-luminous 
infrared galaxies at $z\sim1$ (\cite{pea01}, \cite{takeu01}). These infrared galaxies may be the 
main contributors to the cosmic infrared background radiation (\cite{takeu01}, \cite{chael01}). 
However, the area of the ISOCAM surveys was not large enough to provide sufficient 
samples of these galaxies when separated into several redshift bins, and the surveys 
were made solely with the 15~$\mu$m band. Hence our knowledge on the star-formation 
history around $z\sim1$ is still rather limited.

Thanks to the innovation of the detector technology, the sensitivity in the near and mid 
infrared of AKARI and {\it Spitzer} has been improved by an order of magnitude over 
that of ISOCAM onboard ISO, although the telescope 
aperture has not increased so much in tandem.

\subsection{Spitzer Space Telescope}\label{subsec:comp}

In August 2003 NASA successfully launched  an advanced space infrared observatory, {\it
Spitzer} (\cite{wer04}). {\it Spitzer} is an 85cm diameter
cooled telescope with three 
focal plane instruments, IRAC, MIPS, and IRS. The IRAC instrument covers the near to mid 
infrared wavelengths in 4 bands at 3.6, 4.5, 5.8, 8.0~$\mu$m \citep{faz04}, while 
the MIPS instrument covers the mid to far infrared wavelengths in 3 bands at 24, 70,
160~$\mu$m \citep{rie04}. The IRS instrument \citep{hou04} is a low to moderate resolution
spectrometer covering 5.2-38~$\mu$m with peak-up imagers at 16 and 22~$\mu$m (1.2arcmin$^{2}$ FOVs).

 {\it Spitzer} has already completed various wide area surveys, partly for its {\it Legacy}
science projects (SWIRE, GOODS) and partly for the guaranteed time observations. Data
products from the first look survey (FLS) and the 
{\it Legacy} projects have already been released to the world, and numerous scientific
results have already appeared in the literature. 
Therefore, we should take account the success of {\it Spitzer} in the design of the
AKARI surveys. Besides the major uniqueness as an all-sky surveyor, AKARI also has several unique characteristics  in the 
pointing mode compared to {\it Spitzer}:

\begin{itemize}
\item The 2-26~$\mu$m imaging FOV is four times larger in area (approximately $10'\times10'$) than IRAC, 
\item Continuous wavelength coverage from 2 to 26~$\mu$m, especially in three 
	wavebands (S11, L15, L18W) are available in the 8-24~$\mu$m gap of the {\it Spitzer} bands
\item 2.5-5~$\mu$m slit spectroscopic capability with moderate resolving power of 140
\item 2-26~$\mu$m slitless spectroscopic capability with low resolving power of 20-50
\item four wavebands at 50-180~$\mu$m where {\it Spitzer}/MIPS has two bands.
\end{itemize}

Here we describe the uniqueness of imaging capability in more details.

{\bf 1. FOV of IRC}: The large FOV of the IRC compensates for the lower sensitivity per unit
observing time. Hence the IRC has similar survey speed to that of {\it Spitzer}. However AKARI
has to carry out the all-sky survey and its mission lifetime is shorter than that of 
{\it Spitzer}. Hence, in general {\it Spitzer} as a mission is superior to make deep as well as 
wide-area surveys (of order 10's square degrees).  Note that the IRC-NIR channel can work even in Phase-3 after Helium boil-off. 
Thus, for example, with 10 pointing observations per day over a year in Phase-3, a 50~deg$^2$ area to the depth of ``NEP-Wide" can be surveyed at 2-5~$\mu$m.

{\bf 2. Continuous wavelength coverage from 2-24~$\mu$m}: the IRC wavelength coverage
over the 8-24~$\mu$m gap of the {\it Spitzer} bands (see Figure~\ref{specres}) is extremely useful 
to estimate the redshift and to 
unveil the nature of $z=1-3$ galaxies: star-forming galaxies are characterized 
by the PAH features at rest-frame 3.3, 6.2, 7.7~$\mu$m and silicate absorption at 10~$\mu$m
 (``Silicate-break", see \cite{takpea05}), 
while AGN with dusty torii are characterized by a rather featureless hot dust continuum, 
occasionally with absorption features of silicate or carbon dust grains. 
SED fitting to the sources will break any degeneracy in the starburst-AGN phenomenon.
Moreover, 11-18~$\mu$m surveys which are much deeper and wider than those with ISO are one 
of the most unique capabilities of AKARI. Although {\it Spitzer}/IRS has a 16~$\mu$m peak-up 
camera, it covers only 1.2~arcmin$^{2}$ FOV. Hence, the IRC can 
perform surveys with 100 times faster mapping speed. 

%
 
{\bf 3. Four wavebands at 50-180~$\mu$m}: in the far infrared {\it Spitzer}/MIPS has 70~$\mu$m
 and 160~$\mu$m bands, while AKARI/FIS has 4 wavebands. Such multi-color information
provided by AKARI/FIS is useful to solve the redshift - temperature degeneracy in interpreting
the far infrared SED, since the peak of the dust emission resides in the far-infrared.
This is also true for the studies of the CIRB fluctuations : by cross-correlating the 
faint detected sources or fluctuation images at shorter wavelengths (WIDE-S and N60) 
and the fluctuation images at longer wavelengths (WIDE-L and N160), we can investigate
the typical redshift of the sources contributing the CIRB fluctuations.

By utilizing these unique capabilities, ecliptic pole surveys with AKARI will reveal the dusty star-formation history of the universe, break
the degeneracy in the starburst-AGN phenomenon, and unveil the origin of CIRB.
Regarding the mid-infrared deep survey with AKARI, one might ask why we do not plan to 
observe the {\it Spitzer}/Legacy or FLS fields with solely S11, L15, and L18W. The answer is quite
simple: for deep surveys in the pointing mode, except for regions toward ecliptic poles 
we cannot provide a sufficient number of pointing opportunities to achieve the depth required for 
the detection of $z=2-3$ objects, as described in section~\ref{sec:optimize}. 

It is however, worthwhile to consider collaborative surveys with {\it Spitzer} toward the NEP as
well as the SEP low-cirrus regions. For the former, if IRAC 4 band and MIPS 24~$\mu$m 
imaging were carried out, the 
AKARI/IRC could then concentrate on imaging with 10-20~$\mu$m bands and the slitless spectroscopy.
 In case of the latter survey,
besides the FIS slow-scan surveys we cannot undertake $\sim$10~deg$^2$ IRC surveys because 
of the relatively low visibility of the field. Thus, IRAC and MIPS shallow, but wide-area imaging
will greatly increase the scientific value of the SEP low-cirrus region survey. 

%
%

\section{Coordinated multi-wavelength pre-surveys}\label{sec:ground}

  To perform our scientific goals, multi-wavelength data in the AKARI
deep survey fields at both ecliptic poles are essential in many respects. First of all, 
the higher spatial resolution of ground-based large telescopes enables us to obtain
accurate source positions which are vital for follow-up with optical/near-infrared
spectroscopy. Second, high spatial resolution near infrared images can give us the opportunity to 
classify the galaxy type based on the optical (in rest frame) morphology of the 
source. Third, multi-colour optical/near-infrared photometric data are useful
to estimate the photometric redshift of the counterparts of the AKARI sources
by identifying the Lyman-break (LBGs, \cite{ste96}, \cite{ouc04}), or 
the 0.4~$\mu$m break in their SEDs (such as EROs, \cite{poma00}, \cite{Man02}, \cite{capu04},
\cite{font04}). These techniques are very useful 
to check and confirm the photometric redshifts estimated from the SEDs using the 
AKARI wavebands, by using the 1.6~$\mu$m bump as well as the silicate/PAH features.
We have already obtained preparatory images at the optical/near-infrared bands toward
the AKARI survey fields, and here we briefly introduce the current status for future
reference.

\subsection{NEP pre-surveys}

In June 2003, we observed the ``NEP-Deep" region with the Subaru/Suprime-cam,
where the 3$\sigma$ limiting magnitudes(AB) reach down to $B$=28.4, $V$=27, $R$=27.4,
$i'$=27, and $z'$=26.2 over 918~arcmin$^2$, This has been followed by a
successive observation in July 2004 with $V$ and a narrow band filter (NB711).
Further near-infrared ($J$ and $K_{\rm s}$) images have been taken with 
KPNO-2m/FLAMINGOS to a limiting magnitude (Vega) of $K_{\rm s} \sim 20$ mag also in
June 2004. Together with the near infrared images, we can identify candidate
 active star-forming galaxies at $z > 1.4$ by using the $BzK$ technique
\citep{dad04}. Comparison of these $BzK$ samples with ULIRGs found by
the proposed AKARI deep survey will provide a unique database for the study 
of the star formation history of the Universe. 
 Furthermore, a larger area of approximately 2~deg$^2$ has been
observed with CFHT/Megacam at optical ($g'r'i'z'$) wavelengths to a 5$\sigma$ limiting
magnitude(AB) of 26~mag in the $g'r'i'$ bands and 24~mag in the $z'$ band. This survey covers a
part of the ``NEP-Wide" area and will also be useful in identifying candidates for clusters of
galaxies. We also plan to obtain a deep $U$-band images to search for the Lyman-break population at $z \leq 3$.

\subsection{SEP pre-surveys}

To identify the far infrared sources detected in the FIS slow-scan survey toward
the SEP low-cirrus region, optical observations are underway. In
2004 we obtained R-band images over about half of the FIS survey field area
to $R$=25~mag by using the ESO 2m and the CTIO 4m telescopes. Surveys at other 
wavebands are also under consideration. 

\subsection{Surveys at other wavelengths}

Deep radio images are very useful to identify luminous starburst galaxies.
At the NEP, there is an existing 20~cm VLA observation by \citet{Koll94}
 to a limiting flux density of 200~$\mu$Jy over 29.3~deg$^2$. However,
this survey seems to be too shallow to identify the mid infrared sources 
expected from the deep surveys with AKARI/IRC. For the area centered at the NEP we have 
already take 21~cm radio images to a limiting flux of 50~$\mu$Jy by using the WSRT in the
Netherlands. A deeper VLA proposal is under consideration.

Submillimeter data are mandatory for the study of the star formation history of 
the Universe hidden by the dust, since the submillimeter flux (far-infrared in
rest frame) is a direct measure of the infrared bolometric luminosity originating
from the dust warmed by the stellar light. The SCUBA camera on the JCMT is no longer
available for guest observers and thus 
allocation of the telescope time of the JCMT to the NEP field is very difficult until the
commission of the SCUBA-2 camera.
 The Balloon-borne Large Area Submillimeter Telescope (BLAST)\footnote{see http://chile1.physics.upenn.edu/blastpublic/index.shtml}
 is a promising candidate to obtain submillimeter data for the NEP field. BLAST is a 2-m
aperture balloon-borne telescope incorporating 250, 350, and 500~$\mu$m imagers
with $6.5' \times 13'$ FOV (PI : Dr. Mark Devlin, Univ. of Pennsylvania).
Long duration ($\sim$10~days) flights at the South Pole and Alaska have been comissioned from 2005 summer,
 which will allow 1~deg$^2$ deep surveys to be made. 
We are planning a coordinated multi-band survey toward NEP with AKARI and BLAST.
With this coordinated survey, we will obtain optical identifications of BLAST 
sources with multi-band data in a similar manner to that of the SCUBA galaxies 
with {\it Spitzer}/MIPS 24~$\mu$m fluxes (\cite{ega04}, \cite{ivis04}). 

UV and X-ray data are also very important. 
One of the most promising methods for the selection of $1.5 < z <2.5$ galaxies is the
Lyman break technique which requires near UV observations. Moreover, FUV
fluxes at $\sim 1500$~A are one of the key diagnostics for star formation
while the infrared data from AKARI tell us the amount of starlight 
extincted by the dust. We should seek deep pointing observations with
the Galaxy Evolution Explorer (GALEX)\footnote{see http://www.srl.caltech.edu/galex},
which was launched in April 2003. 
X-ray data are also important in two respects:  the X-ray emission
from the hot intergalactic medium is useful to search for cluster candidates,
and deep X-ray images are necessary to identify buried AGN which
may be strong mid-infrared emitters in the AKARI deep images.
The ROSAT source catalogues have already been released
(\cite{hen01}, \cite{gio03}), and hard X-ray imaging with 
XMM-Newton is under consideration.

\section{Summary}\label{sec:summary} 

The AKARI observational strategy and the details of the observational plans 
for the coordinated deep pointing surveys toward the ecliptic poles are described.
We have reviewed the scientific goals and requirements for the surveys, as well as the 
technical constraints of the AKARI spacecraft and effects of the foreground
astronomical sources. We conclude that the NEP is the best location for the AKARI
extragalactic 2-26~$\mu$m deep survey, while the low-cirrus regions near
SEP is optimum for the 50-180~$\mu$m slow-scan surveys. The ELAIS-N1 field, one of 
the {\it Spitzer}/SWIRE fields, cannot be selected as a wide, shallow survey field 
since its ecliptic latitude is not high enough to generate a circular or nearly 
rectangular map.

The technical details of the blank-field surveys are as follows:
\begin{itemize}
\item ``NEP-Deep" 2-26~$\mu$m survey: deep (28 pointing observations per field-of-view)
  survey of approximately 0.5~deg$^{2}$ with three filters per each IRC channel,
  resulting in 504 pointing observations in total(a half is in Phase-1), and
\item ``NEP-Wide" 2-26~$\mu$m survey: wide but shallow (2 pointing per FOV) survey 
  of approximately 6.2~deg$^{2}$ with three filters per each IRC channel. 
  The survey region is a circular area surrounding the `` NEP-Deep" field, 
  and in total 450 pointing observations are required.
\item SEP low-cirrus region survey at 50-180~$\mu$m : 15-20~deg$^2$ fan-shape area survey 
  toward the very low cirrus ($\bar{B}_{100}$ is as low as 0.2MJy/sr) region by using
 the FIS slow-scan mode with a redundancy of at least two. In total 270 pointing 
 observations are required.
\end{itemize}

In the same target field of the above imaging surveys, we are also investigating the
technical feasibility of IRC's unique slitless spectroscopic 
capability, in order to construct an unbiased sample of near to far infrared SED 
templates for various kinds of galaxies at $z<1$. This will be further described
in a separate paper (Ohyama et al. in prep.). Multiwavelength ground-based surveys
are also ongoing, and optical images and catalogs will be published soon (Wada et al.
in prep.).

\section*{Acknowledgements} 

We would like to thank all AKARI team members for their support on this 
project. We also thank Dr. David Hughes for discussion on possible collaboration
between BLAST and AKARI,  Dr. Takamitsu Miyaji for his effort on the X-ray observation plan, and Dr. Patrick Henry for information on the ROSAT clusters toward NEP. This work 
partly supported by the JSPS grants (grant number 15204013, and 16204013). CPP acknowledges support from JSPS while in Japan.

\end{document}